\documentclass[aos,preprint]{imsart}

\RequirePackage[OT1]{fontenc}
\RequirePackage{amsthm,amsmath}
\RequirePackage[numbers]{natbib}
\usepackage{amssymb,cmmib57,graphics,color,mathrsfs,eufrak,dsfont}
\usepackage{graphicx}
\usepackage{xcolor,soul}
\usepackage{pst-plot}
\usepackage{url}
\usepackage{algorithm,algpseudocode}
\usepackage{multirow}
\newcommand{\bbzero}{{\bf 0}}
\def\ep{\varepsilon}
\newcommand{\bX}{\mathbf{X}}

\newcommand{\bA}{{\mbox{\boldmath$A$}}}
\newcommand{\bW}{{\mbox{\boldmath$W$}}}

\newcommand{\ma}{\mathcal{A}}

\newcommand{\bP}{{\mbox{\boldmath$P$}}}

\newcommand{\mM}{\mathcal{M}}

\arxiv{arXiv:0000.0000}

\startlocaldefs
\numberwithin{equation}{section}
\theoremstyle{plain}
\newtheorem{thm}{Theorem}[section]
\endlocaldefs

\begin{document}

\begin{frontmatter}
\title{Multi-threshold Accelerated Failure Time Model\thanksref{T1}}
\runtitle{Multi-threshold AFT Model}
\thankstext{T1}{Supported by National Medical Research Council NMRC/CBRG/0014/2012 in Singapore, Academic Research Funding R-155-000-174-114  and National Natural Science Foundation of China (General Program, No. 11571337; Key Program, No. 71631006).}

\begin{aug}
\author{\fnms{Jialiang} \snm{Li}\thanksref{m2}\ead[label=e2]{stalj@nus.edu.sg}}
\and
\author{\fnms{Baisuo} \snm{Jin}\thanksref{m1}\ead[label=e1]{jbs@ustc.edu.cn}}
%\thankstext{t1}{Some comment}
%\thankstext{t2}{First supporter of the project}
%\thankstext{t3}{Second supporter of the project}
\runauthor{J. Li and B. Jin}

\affiliation{National University of Singapore\thanksmark{m2} and University of Science and Technology of China\thanksmark{m1} }

\address{
Department of Statistics and Applied Probability\\
National University of Singapore\\
Singapore,119077\\
\phantom{E-mail:\ }\printead{e2}}

\address{
Department of Statistics and Finance \\
University of Science and Technology of China\\
Hefei, Anhui, China, 230026\\
\phantom{E-mail:\ }\printead{e1}}

\end{aug}

\begin{abstract}
A two-stage procedure for simultaneously detecting multiple
thresholds and achieving model selection in the segmented accelerated failure time (AFT) model is developed in this paper.
In the first stage, we formulate the threshold problem as a group model selection problem so that a concave 2-norm group  selection method can be applied. In the second stage, the thresholds are finalized via a refining method. We establish the strong consistency  of the threshold estimates and regression coefficient estimates under some mild technical conditions. The proposed procedure performs satisfactorily in our simulation studies. Its real world applicability is demonstrated via analyzing a follicular lymphoma data.
\end{abstract}



\end{frontmatter}
 %\linespread{2}

\section{Introduction}
Applied economists routinely test their models for the presence of structural change. If the evidence supports it, a threshold model is constructed and one needs to detect the thresholds (also called break points) at which to split the sample. The threshold variable may be an element of regressors. One such example is the well-known threshold autoregressive  model (see Tong, 2012).
Sometimes the threshold variable is simply the index of observed samples (eg. time in a time series model) and the model is commonly referred to as the change point model or segmented regression model (see Yao and Au, 1989, Perron, 2006, Fearnhead and Vasileiou, 2009).

In this paper we focus on the setting of multiple thresholds which is a much more challenging problem than a single break-point detection (Hansen, 2000). A number of issues arise in the presence of multiple change points. These include
the determination of the number of breaks, estimation of the thresholds
given the number, and statistical analysis of the resulting estimators. There exists a rich literature on this subject. For example,
 Inclan and Tiao (1994) identified  multiple change-points of variance using the iterated cumulative sum of squares (ICSS) algorithm. Bai and Perron (2003)  developed  the dynamic programming principle  for the estimation of multiple change-points in linear regression. Following the familiar idea of penalized estimation, Harchaoui and L\'{e}vy-Leduc (2010) used the least absolute shrinkage and selection operator (LASSO) algorithm to estimate the locations of multiple change-points, in one-dimensional piecewise constant signals. Davis et al. (2006) proposed a genetic algorithm to detect multiple break points, while recently Jin, Shi and Wu (2013) considered non-concave penalty functions including the smoothly clipped absolute deviation (SCAD) penalty and minimax concave penalty (MCP) penalty in piecewise stationary autoregressive processes. {  However, computational procedure and theoretical justification in Jin, Shi and Wu (2013) cannot be easily extended to survival analysis. We will consider a more difficult setting with censored event time in this paper. The theoretical results established in this paper are also more general than those in Jin, Shi and Wu (2013).}

{  In life-testing research studies, single change point problem has been addressed by many authors. Specifically, Luo, Turnbull and Clark (1997) considered the Cox model with a change point at an unknown time and established asymptotic results for maximum partial likelihood estimates. Pons (2003) studied unknown threshold of a predictor variable under Cox model using counting processes theory while Kosorok and Song (2007) further examined the more sophisticated linear transformation models and provided necessary inference tools.} Very few authors considered the problem of estimating multiple thresholds  for survival regression analysis. Censored lifetime data break down the usual estimation framework for completely observed data. We usually cannot attain a closed-form solution for regression estimates and therefore face an increased complexity of estimating thresholds accurately. To address this under-developed issue, we consider the accelerated failure time (AFT) model as a typical example of regression models in this paper and contribute a new methodology on change-point problem for survival data analysis.

The AFT model permits a direct assessment of the covariate effects on the survival time, facilitating the interpretation of regression coefficients for the mean response. There are many estimation methods available for AFT model in the literature, including Buckley and James (1979), Prentice (1978), Tsiatis (1990), Ying (1993), Lin, Wei and Ying (1998). Many estimation methods for right censored data are rank-based and in practice the estimating functions may be discontinuous, producing challenges to the computation. In contrast, Stute (1993, 1996) proposed a weighted least squares estimator for AFT model, and established the consistency and asymptotic normality under technical conditions. Huang, Ma and Xie (2006) carried out a variable selection procedure and estimation in AFT model with high-dimensional covariates based on Stute's estimator. Xia, Jiang, Li and Zhang (2016) also employed Stute's estimator for nonparametric variable screening and selection. In this paper, we formally adopt the Stute estimator to study the parameter estimation for the AFT model with $s$ thresholds, where $s\geq 0$ is unspecified. Using a two-stage procedure proposed in this paper we may estimate the thresholds and the regression coefficients simultaneously. Interestingly our procedure may be straight forwardly extended to incorporate variable selection for high-dimensional data analysis.

The paper is arranged as follows. In Section 2, the multiple break-point problem for the AFT model is formulated. We then propose a two-stage procedure to detect the thresholds and estimate model parameters. In Section 3, we establish theoretical properties of our procedure. Next simulation studies are
conducted in Section 4 to examine the performance of our methods. An empirical application to a follicular lymphomais data is presented in Section 5.

Throughout the paper, ${\mathbf 1}_q=(1,\ldots,1)^{\top}  $ is a
$q$-dimensional constant vector, $I_q$ is the $q\times q$ identity matrix,  $1_{\{\cdot\}}$ is an
indicator function, $\bA^{\top}  $ is the transpose of a
matrix $\bA$, and $\lfloor c\rfloor$ is the
integer part of a real number $c$. For a vector $\mathbf  a$, $\mathbf  a^{\top}  $ is
its transpose, $a_j$ is its $j$th component, $|\mathbf  a|$ and  $\|\mathbf  a\|$
 are respectively its $L_1$-norm and $L_2$-norm. If $\ma$ is a set, its
complement and its size are denoted by ${\ma}^c$ and $\sharp\ma$,
respectively. In addition, ``$\rightarrow_{a.s.}$'' denotes convergence with probability 1.

\section{Methodology for multiple thresholds under AFT model}

Let $T_i,i=1,\ldots,n$ be the independent logarithm of the failure time and $\mathbf X_1,\ldots,\mathbf X_n$ are i.i.d.  $p$-dimensional regressors.
Assume $(T_i,\mathbf X_i), i=1,\ldots, n$, satisfy the following AFT model with $s$
thresholds located at $a_{1}<\ldots<a_{s}$:
{\small\begin{eqnarray}
T_i&=&\sum_{j=1}^{s+1}\mathbf X_i^{\top}  \mathbf \beta_{j}^*1_{\{a_{j-1}<Z_i\leq a_{j}\}}+\varepsilon_{i}\nonumber\\
&=&\mathbf X_i^{\top}  \left[\mathbf \beta_{1}^*+\sum_{\ell=1}^{s}\mathbf d_{\ell}^*
1_{\{a_{\ell}< Z_i\leqslant a_{s+1}\}}\right]+\varepsilon_{i},\quad
i=1,\ldots, n\nonumber\\
&=&{\begin{cases}\mathbf X_i^{\top}  \mathbf \beta_{1}^*+\varepsilon_{i} &\mbox{if } a_{0}< Z_i\leq a_{1}, \\
\mathbf X_i^{\top}  (\mathbf \beta_{1}^*+\mathbf d_{1}^*)+\varepsilon_{i} & \mbox{if } a_{1}<Z_i\leq a_{2}, \\
\ldots&\ldots\\
\mathbf X_i^{\top}  \left(\mathbf \beta_{1}^*+\sum_{\ell=1}^{s}\mathbf d_{\ell}^*\right)+\varepsilon_{i}, & \mbox{if } a_{s}<Z_i\leq a_{s+1}, \end{cases}}\label{cp}
\end{eqnarray}
}
where $\mathbf \beta_1^*, \ldots, \mathbf \beta_{s+1}^*$ are  unknown $p$-dimensional regression coefficients for $s+1$ subgroups, $\mathbf  d_{\ell}^*=\mathbf \beta_{\ell+1}^*-\mathbf \beta_{\ell}^*,\ell=1,\ldots,s$, are the increments of coefficients between two adjacent subgroups, $s\geq 0$ is the unknown number of thresholds,
$Z_i$ is the thresholding variable,
$a_{1}, \ldots, a_{s}$ are unknown
threshold locations, $a_{0}=0$, $a_{s+1}=n$,
 $\varepsilon_1, \ldots, \varepsilon_n$ are independent random errors, and
$\varepsilon_{a_{\ell-1}+1},\ldots,\varepsilon_{a_{\ell}}$ , {  the subgroup in the series $\{\varepsilon_i,i=1,\ldots,n\}$ separated by the threshold locations $\{a_\ell, \ell=1,\ldots, s\}$}, are identically distributed with mean zero and variance $\sigma_{\ell}^2$, $\ell=1,\ldots,s+1$.
Notice that if $s=0$, the model (\ref{cp}) does not involve threshold. If $s\geq 1$, we denote $P(Z_i\leq a_j)=\tau_j,j=1,\ldots,s$ where
$0<\tau_1<\tau_2<\ldots<\tau_s<1$. If $Z_i=i$, we retain the usual change point model, and the condition $P(Z_i\leq a_j)=\tau_j$ will be replaced by $a_j/n\to \tau_j$ since change point locations actually depend on the sample size in this setting. Figure 1 is an example of model (\ref{cp}), where $s=2$, $Z$ is generated from a normal distribution and $T$ is generated from a mixture normal distribution.

%\newgeometry{left=3.3cm,right=3.3cm,top=1cm}
\begin{figure}[htb!]\label{fig0}
\begin{center}
{
%\centerline
{
\includegraphics[bb=0 0 722 399,scale=0.35]{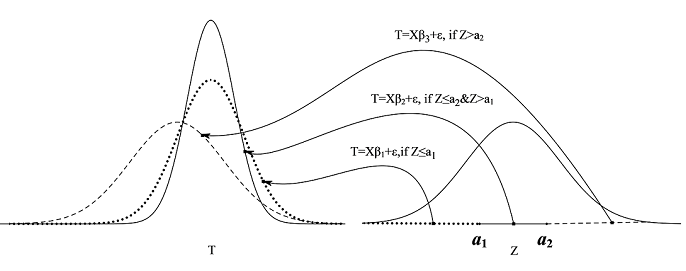}
}}
\caption{An example of model (\ref{cp})}
\end{center}
\end{figure}

%\restoregeometry

Let $C_i,i=1,\ldots,n$ be logarithm of the censoring time which are assumed to be independent and identically distributed (i.i.d.). In practice we only observe $(Y_i,\delta_i,\mathbf X_i),i=1,\ldots,n$, where $Y_i=\min(T_i,C_i)$ is the  censored logarithm of the failure time and $\delta_i=1_{\{T_i\leq C_i\}}$
is the censoring indicator.

Throughout this paper,  let $Y_{(1,\mathcal{I})}\leq \ldots \leq Y_{(b,\mathcal{I})}$ be the ordered failure times in the index set $\mathcal{I}$  and $b=\sharp\mathcal{I}$. Let $\delta_{[\ell,\mathcal{I}]}$ be the
concomitant of the $\ell$th-ordered $Y_{(\ell,\mathcal{I})}, \ell=1,\ldots,b$. Define the Kaplan-Meier weights as
{ $$w_{[1,\mathcal{I}]}=\frac{\delta_{[1,\mathcal{I}]}}{b}, w_{[\ell,\mathcal{I}]}=\frac{\delta_{[\ell,\mathcal{I}]}}{b-\ell+1}\prod_{k=1}^{\ell-1}
\big(\frac{b-k}
{b-k+1}\big)^{\delta_{[k,\mathcal{I}]}},\qquad \ell=2,\ldots, b.$$}
Such weights are constructive elements of the Kaplan-Meier estimator for the survival function (See Lawless, 2011).
Let $r_{Y_i,\mathcal{I}}\in [1,b]$ be the rank of $Y_i$ among $\{Y_{i}: i\in \mathcal{I}\}$  i.e. $Y_i=Y_{(r_{Y_i,\mathcal{I}},\mathcal{I})}$, $i\in \mathcal{I}$. For simplicity, we denote $Y_{(r_{Y_i,\mathcal{I}})}=Y_{(r_{Y_i,\mathcal{I}},\mathcal{I})}=Y_i$ and $w_{[r_{Y_i,\mathcal{I}}]}=w_{[r_{Y_i,\mathcal{I}},\mathcal{I}]}$.

If $a_{j}, j=1,\ldots,s$ are known, define $\mathcal{I}_j^*=\{i:a_{j-1}<Z_i\leq a_{j}\}$ and $b_j^*=\sharp \mathcal{I}_j^*$, we can use the Stute estimator to fit the AFT model. The resulting estimator $\mathbf {\hat \beta}_1^*, \mathbf {\hat d}_1^*, \ldots, \mathbf {\hat d}_s^*$ may be represented by a weighted least squares estimator
 that
minimizes
\begin{eqnarray}\label{ls}
\sum_{j=1}^{s+1}\frac{b_j^*}{2n}\sum_{i\in \mathcal{I}_j^*}
w_{[r_{Y_i,\mathcal{I}_j^*}]}\left(Y_{i}-\mathbf X_i^{\top}  (\mathbf \beta_1^*+\sum_{k=1}^{j-1}\mathbf d_k^*)\right)^2.
\end{eqnarray}
We note that the weights $\{w_{[r_{Y_i,\mathcal{I}_j^*}]}\}$ depend on the order of the random failure times and hence the related asymptotic argument is not as standard as the independent and identically distributed (i.i.d.) case in linear regression.

We intend to estimate $s$, $a_{1}$, $\ldots$, and
$a_{s}$ and then use the detected change point(s) to obtain the Stute estimator for regression coefficients. To this end, first we need to decide the total number of thresholds and the approximate distances between pairs of thresholds. After that we may proceed to determine the exact locations of thresholds.  We propose a two-stage procedure in the following sections.

\subsection{The Splitting Stage}

First we split the data sequence into $q_n+1$ segments where $q_n$ tends to infinity as $n\to \infty$. Let $\{Z_i:\delta_i=1,i=1,\ldots,n\}=\{\tilde Z_1,\ldots,\tilde Z_{n^*}\}$ where $n^*=\sum_{i=1}^n\delta_i$ is the total number of events. In order to allow each segment to include enough failure event observations, the data sequence is split such that the first segment $\mathcal{I}_{1}=\{i: Z_i\leq \tilde Z_{(n^*-q_nm)}\}$ involves
 $n^*-q_nm$ events, and each of the other $q_n$ segments  $\mathcal{I}_{j}=\{i: \tilde Z_{(n^*-(q_n-j+2)m)}<Z_i\leq \tilde Z_{(n^*-(q_n-j+1)m)}\},j=2,\ldots,q_n+1$ involve
$m$ events where $m=\lceil n^*/q_n\rceil$ and $\tilde Z_{(1)}\leq \tilde Z_{(2)}\ldots\leq \tilde Z_{(n^*)}$ are the ordered threshold variables associated with subjects no failed.
Define $b_j=\sharp\mathcal{I}_{j}, j=1, \ldots, q_n+1$.

To select these intervals covering thresholds $\{a_{j}\}$, we propose a concave 2-norm group selection method, such that an estimate $\hat {\mathbf \theta}=(\hat{\mathbf \beta}_1^{\top}  ,\hat{\mathbf d}_1^{\top}  ,\ldots,\hat{\mathbf d}_{q_n}^{\top}   )^{\top}  $  is given by minimizing
\begin{align}\label{esti0}
\sum_{j=1}^{q_n+1}\frac{b_j}{2n}\sum_{i\in \mathcal{I}_{j}}
w_{[r_{Y_i,\mathcal{I}_{j}}]}\left(Y_{i}-\mathbf X_i^{\top}  (\mathbf \beta_1+\sum_{k=1}^{j-1}\mathbf d_k)\right)^2+ \sum_{k=1}^{q_n}
p_{\lambda_n,\gamma_n}( \|\mathbf d_{k}\|),
\end{align}
where $\lambda_n>0$, $\gamma_n>1$ are tuning parameters, and the penalty function $p_{\lambda_n,\gamma_n}( |u|)>0$  is concave in $|u|$.
For the simplicity of presentation, we shall only consider two well-studied penalty functions in this paper, namely the smoothly clipped absolute deviation (SCAD) penalty and the minimax concave penalty (MCP). SCAD was introduced in Fan and Li (2001) and defined as
$p_{\lambda,\gamma}(u)=\lambda uI_{[0,\; \lambda]}(u)+\frac{\gamma\lambda
u-0.5(u^2+\lambda^2)}{\gamma-1}I_{(\lambda,\;\gamma\lambda]}(u) +
\frac{\lambda^2(\gamma^2-1)}{2(\gamma-1)}I_{(\gamma\lambda,\;\infty)}(u),\gamma>1.$  MCP was introduced in Zhang (2010) and defined as
$ p_{\lambda,\gamma}(u)= (\lambda u-\displaystyle\frac{u^2}{2\gamma})I_{[0, \;\gamma
\lambda]}(u)+\displaystyle\frac{1}{2} \gamma\lambda^2I_{(\gamma\lambda,\;\infty)}(u),\gamma>2.
$

Let $\mathbf Y_{(j)}=(Y_{i},i\in \mathcal{I}_j)^{\top}  $, $\bX_{(j)}=(\mathbf X_i,i\in \mathcal{I}_j)^{\top}  $,
$\tilde {\mathbf w}_{(j)}=b_j(w_{[r_{Y_{i},\mathcal{I}_j}] },i\in \mathcal{I}_j)^{\top}  $,
$\tilde {\mathbf y}_{(j)}=diag(\tilde  {\mathbf w}_{(j)})^{1/2}\mathbf Y_{(j)}$, $\tilde \bX_{(j)}=diag(\tilde  {\mathbf w}_{(j)})^{1/2}\bX_{(j)}$, $j=1,\ldots,q_n+1$. Denote
$\tilde {\mathbf y}=(\tilde {\mathbf y}_{(1)}^{\top}  ,\ldots,\tilde {\mathbf y}_{(q_n+1)}^{\top}  )^{\top}  $,
$\tilde \bX=(\tilde \bX^{(1)},\ldots,\tilde \bX^{(q_n+1)})$ where $\tilde \bX^{(1)}=(\tilde \bX_{(1)}^{\top}  ,\ldots,\tilde \bX_{(q_n+1)}^{\top}  )^{\top}  $ and
$\tilde \bX^{(j)}=(\bbzero_{p\times \sum_{i=1}^{j-1}b_i}, \tilde \bX_{(j)}^{\top}  $, $\ldots$, $\tilde \bX_{(q_n+1)}^{\top}  )^{\top}  $, $j=2, \ldots$, $q_n+1$.
The estimator $\hat{\mathbf \theta}$ in (\ref{esti0}) can be written as
\begin{eqnarray}\label{esti}
\hat{\mathbf \theta}=\arg\min_{\mathbf \theta} \left\{\frac{1}{2n}\| \tilde {\mathbf y}- \tilde \bX \mathbf \theta \|^2+ \sum_{j=1}^{q_n}
p_{\lambda_n,\gamma_n}( \|\mathbf d_{j}\|)\right\},
\end{eqnarray}
We apply the group coordinate descent (GCD) algorithm to estimate $\hat{\theta}_n$ from (\ref{esti}).

The properties of $\hat{\theta}$
will be given by Theorem \ref{thm2} in the next section. For simplicity, we write $\hat{\mathbf \theta}=(\hat{\mathbf \theta}_1^{\top}  ,
\ldots,\hat{\mathbf \theta}_{q_n+1}^{\top}  )^{\top}  $ such that $\hat{\mathbf \theta}_1=\hat{\mathbf \beta}_1$ and  $\hat{\mathbf \theta}_j=\hat{\mathbf  d}_{j-1}$, $j=2,\ldots,q_n+1$.
Let $\hat \ma=\{j:\hat {\mathbf \theta}_j\neq 0, j=1,\ldots,q_n+1\}$, and  $\hat \ma^*$ be a subset of $\hat \ma$ such that
$\hat \ma^*=\{j:j\in \hat \ma,j-1\not \in \hat \ma,j=2,\ldots,q_n+1\}=\{\hat k_{1},\ldots, \hat k_{\hat s}\}$. If $\hat s=0$, we declare there is no threshold.
If $\hat s>0$, by Theorem \ref{thm2}, the true threshold $a_{j}$ is highly likely to be located in $
( \tilde Z_{(n^*-(q_n-\hat k_{j}+3)m)}, \tilde Z_{(n^*-(q_n-\hat k_{j}+1)m)}]$, $j=1,\ldots,\hat s$.

\subsection{The Refining Stage}

If $\hat s>0$, by Theorem \ref{thm2}, we can estimate the threshold $a_{j}$ in $(\tilde Z_{(n^*-(q_n-\hat k_{j}+3)m)}$, $\tilde Z_{(n^*-(q_n-\hat k_{j}+1)m)}]$. To obtain the  estimates of the thresholds $\{a_{j}\}$, we
denote $\hat{\mathcal{I}}_j=\{i:\tilde Z_{(n^*-(q_n-\hat k_{j}+3)m)}<Z_i \leq \tilde Z_{(n^*-(q_n-\hat k_{j}+1)m)}\}$,
$\hat{\mathcal{I}}_{j,\varsigma^-}=\{i:\tilde Z_{(n^*-(q_n-\hat k_{j}+3)m)}<Z_i \leq \varsigma\}$, $\hat{\mathcal{I}}_{j,\varsigma^+}=\{i:\varsigma<Z_i \leq \tilde Z_{(n^*-(q_n-\hat k_{j}+1)m)}\}$
\begin{eqnarray*}Q_{j}(\varsigma^-,\mathbf \beta)&=&  \frac{\sharp \hat{\mathcal{I}}_{j,\varsigma^-}}{\sharp \hat{\mathcal{I}}_{j}}\sum_{i\in \hat{\mathcal{I}}_{j,\varsigma^-}}w_{[r_{Y_i,\hat{\mathcal{I}}_{j,\varsigma^-}}]}(Y_{i}-\mathbf X_{i}^{\top}  \mathbf \beta)^2, \\
Q_{j}(\varsigma^+,\mathbf \beta)&=&  \frac{\sharp \hat{\mathcal{I}}_{j,\varsigma^+}}{\sharp \hat{\mathcal{I}}_{j}}\sum_{i\in \hat{\mathcal{I}}_{j,\varsigma^+}}w_{[r_{Y_i,\hat{\mathcal{I}}_{j,\varsigma^+}}]}(Y_{i}-\mathbf X_{i}^{\top}  \mathbf \beta)^2, \end{eqnarray*}
and use the following method to estimate $a_{j}$:
\begin{align}\label{esta}\hat a_{j}=\mbox{argmin}_{\varsigma\in (\tilde Z_{(n^*-(q_n-\hat k_{j}+3)m)},\tilde Z_{(n^*-(q_n-\hat k_{j}+1)m)}]\bigcap\{Z_{(1)},\ldots,Z_{(n)}\} }\{ Q_j(\varsigma)\},\end{align}
where $Q_j(\varsigma)= \min_{\mathbf \beta}Q_{j}(\varsigma^-,\mathbf \beta)+\min_{\mathbf \beta} Q_{j}(\varsigma^+,\mathbf \beta)$, $j=1,\ldots,\hat s$, and
$Z_{(1)}\leq Z_{(2)}\ldots\leq Z_{(n)}$ are the order of $Z_i,i=1,\ldots,n$
. The regions separated by the thresholds achieve the overall minimum least squares errors. The consistency of $\hat a_{j}$
will be provided in Theorem \ref{thm3}.

\subsection{Remarks}

After we obtain $\hat a_{j},j=1,\ldots,\hat s$ by (\ref{esta}), it is sensible to use the weighted least squares to obtain a final estimate of the coefficients in model (\ref{cp}). Let $\hat{\mathcal{I}}^*_j=\{i: \hat a_{j-1}<Z_i\leq \hat a_{j}\}$ and $\hat b_j^*=\sharp \hat{\mathcal{I}}^*_j$, $j=1,\ldots,\hat s+1$ where $\hat a_{0}=-\infty$ and $\hat a_{\hat s+1}=+\infty$. Then the coefficient $\mathbf \theta^*=((\mathbf \beta^*_1)^{\top}  ,(\mathbf d_1^*)^{\top}  ,\ldots,(\mathbf d_{\hat s}^*)^{\top}  )^{\top}  =
(\theta_1^*,\ldots,\theta_{p(\hat s+1)}^*)^{\top}  $ can be estimated by minimizing the following penalized least squares
\begin{align}\label{estit}
M(\mathbf \theta)=\sum_{j=1}^{\hat s_n+1}\frac{\hat b_j^*}{2n}\sum_{i\in \hat{\mathcal{I}}^*_j }
w_{[r_{Y_i,\hat{\mathcal{I}}^*_j}]}\left(Y_{i}-\mathbf X_i^{\top}  (\mathbf \beta_1+\sum_{k=1}^{j-1}\mathbf d_k)\right)^2+ \sum_{i=1}^{p(\hat s_n+1)}
p_{\lambda_n,\gamma_n}( |\theta_i|),
\end{align}
where the penalty function $p_{\lambda_n,\gamma_n}( |u|)>0$ is the same as in (\ref{esti}) and $\mathbf \theta=(\mathbf \beta_1^{\top}  ,\mathbf d_1^{\top}  ,\ldots,\mathbf d_{\hat s}^{\top}  )^{\top}  =
(\theta_1,\ldots,\theta_{p(\hat s+1)})^{\top}  $. The inclusion of the penalty functions may lead to a sparse solution since the number of parameters could be quite large for all the subgroups. The oracle property of $\hat{\mathbf \theta}^*=\arg\min_{\mathbf \theta}M(\mathbf \theta)$
will be given by Theorem 3.3.

Similar to Jin, Shi and Wu (2013), $\gamma_n$ is set as 2.4 for SCAD and MCP penalties . The regularization parameter $\lambda_n$ can be chosen by the Bayesian information criterion (BIC).

The performance of $\hat{\mathbf \theta}^*$ is further dependent on the segment length $m$. The selection of an optimal $m$ can be carried out as follows: Apply the splitting stage to the data sequence $L$ times with the common segment length of events (excluding the first segment) $m_\ell$, $\ell=1,\ldots,L$. In order to meet the assumption, i.e., $m=\lfloor c\sqrt n \rfloor$ made in Theorem \ref{thm2}, we set $m_\ell=\lfloor\kappa_\ell\sqrt n \rfloor$, $\ell=1,\ldots,L$, where $\kappa_\ell$ takes values from $L$ grid-points in an interval, say [0.1,\;2.0]. For each $m_\ell$, applying the proposed two-stage procedure, we obtain the set of estimated thresholds $\hat \mM_{\ell}=\{\hat a_{1,\ell},\ldots,\hat a_{\hat s_{\ell},\ell}\}$, $\ell=1,\ldots,L$. We use the BIC to choose the best index
\begin{equation}\label{final}\hat \ell=\mbox{argmin}_{\ell=1,\ldots,L}\left\{BIC_{\hat \mM_{\ell}}\right\},\end{equation}
where \begin{align}\label{bic}
BIC_{\hat \mM_{\ell}}=& n\log\left(\sum_{j=1}^{\hat s_{\ell}+1}\frac{\sharp\hat{\mathcal{I}}_{j,\ell}}{n}\left(\min_{\mathbf {\beta}}\sum_{i\in \hat{\mathcal{I}}_{j,\ell} }w_{[r_{Y_i,\hat{\mathcal{I}}_{j,\ell}}]}(Y_i-\mathbf {X}_i^{\top}  \mathbf {\beta})^2\right)\right)\nonumber\\ &+q(\hat s_{\ell}+1)\log(n)
\end{align}
with $\hat{\mathcal{I}}_{j,\ell}=\{i:\hat a_{j-1,\ell}<Z_i\leq \hat a_{j,\ell}\}$, $\hat a_{0,\ell}=-\infty, \hat a_{\hat s_{\ell} +1,\ell}=+\infty$. The optimal ${m}_{opt}=\lfloor\kappa_{\hat\ell}\sqrt n \rfloor$ and the optimal estimated thresholds $\hat \mM_{opt}=\{\hat a_{1, \hat \ell},\ldots,\hat a_{\hat s_{\hat\ell}, \hat\ell}\}$.

We refer to the proposed two-stage procedure as \textit{T}wo \textit{S}tage \textit{M}ultiple \textit{C}hange-points \textit{D}etection (TSMCD) from now on. Since two regularization methods, i.e. MCP, and SCAD, will be utilized in the splitting stage, we refer to the corresponding TSMCD as  TSMCD{\scriptsize{(MCP)}} and TSMCD{\scriptsize{(SCAD)}}, respectively. The detailed algorithm of TSMCP is described in the following table.

\begin{algorithm}
\caption{TSMCD}
\label{alg1}
\begin{algorithmic}[1]
\For{$\ell= 1, 2,\ldots, 20$}

Step 1: Splitting stage
\State {Set $m=\lfloor 0.1 \ell\sqrt {n^*} \rfloor$  and $q_n= \lfloor n^*/m \rfloor-1$ where $n^*$ is the number of events; }
\State {Split the data sequence into $q_n+1$ segments $\mathcal{I}_{j},j=1,\ldots,q_n+1$ as Section 2.1;}
\State {Estimate  $\hat {\mathbf \theta}=((\hat{\mathbf \theta}_1)^{\top}  ,
\ldots,(\hat{\mathbf \theta}_{q_n+1})^{\top}  )^{\top}  $  by minimizing (\ref{esti0}) or (\ref{esti});}
\State {Compute the index sets $\hat \ma=\{j:\hat {\mathbf \theta}_j\neq 0, j=1,\ldots,q_n+1\}$ and
$\hat \ma^*=\{j:j\in \hat \ma,j-1\not \in \hat \ma,j=2,\ldots,q_n+1\}\equiv\{\hat k_{1},\ldots, \hat k_{\hat s}\}$ where $\hat k_{1}<\hat k_{2}<\ldots<\hat k_{\hat s}$ and $\hat s=\sharp \hat \ma^*$; }

Step 2: Refining stage
\If{$\hat s=0$} {go to the step 9;}
\ElsIf{$\hat s>0$} {estimate the threshold $a_j$ in $(\tilde Z_{(n^*-(q_n-\hat k_{j}+3)m)}$, $\tilde Z_{(n^*-(q_n-\hat k_{j}+1)m)}]$  by (\ref{esta});}
\EndIf
\State {Estimate the coefficient $\mathbf \theta^*=((\mathbf \beta^*_1)^{\top}  ,(\mathbf d_1^*)^{\top}  ,\ldots,(\mathbf d_{\hat s}^*)^{\top}  )^{\top}  $ by minimizing (\ref{estit});}
\State {Return the set of estimated thresholds $\hat \mM_{\ell}=\{\hat a_{1,\ell},\ldots,\hat a_{\hat s_{\ell},\ell}\}$ in step 7 and the estimator of the coefficient $\hat{\mathbf {\theta}}^*_{\ell}$ in step 9, and compute $BIC_{\hat \mM_{\ell}}$ by (\ref{bic});  }
\EndFor
\State {Choose $\hat{\ell}$ that minimizes the $BIC_{\hat \mM_{\ell}}$ and obtain the final estimators $\hat \mM_{opt}=\hat \mM_{\hat{\ell}}$ and $\hat{\mathbf {\theta}}^*_{opt}=\hat{\mathbf {\theta}}^*_{\hat{\ell}}$.}
\end{algorithmic}
\end{algorithm}

\par

\bigskip
%%%%%%%%%%%%%%%%%%%%%%%%%%%%%%%%%%%%%%%%%%%%%%%%%%%%%%%%%%%%%%%%%%%%%%%%%%%%%%%%%%%%%%%%%%%%%%%%%%%%%%%%%%%%%%%%%%%%%%%%%%%%
%
%\noindent{\large\bf References}
%\begin{description}
%%\item
%%Andrews, D. W. K. (1984). Non-strong mixing autoregressive processes.
%%  {\it J. Appl. Probab.} {\bf 21}, 930-934.
%%\item
%%Avram, F. and Taqqu, M. S. (1987)  Noncentral limit theorems and Appell
%%polynomials. {\it Ann. Probab.} {\bf 15}, 767-775.
%%\item
%%Bradley, R. C. (1986). Basic properties of strong mixing conditions. In {\it
%%Dependence
%%in Probability and Statistics} (Edited by E. Eberlein and M. S. Taqqu),
%%162-192. Birkh\"auser, Boston.
%%\item
%%Fox, R. and Taqqu, M. (1987). Central limit theorems for quadratic forms in
%%random
%%variables having long-range dependence. {\it Probab. Theory Related Fields}
%%{\bf 74}, 213-240.
%
%
%
%
%\end{description}
\section{Asymptotic theory}
 Under model (\ref{cp}), we further denote the distributions of $\{T_i,i\in \mathcal{I}_j^*\}$ and $\{Y_i,i\in \mathcal{I}_j^*\}$ to be $F_{j}$ and $H_{j}$, $j=1,\ldots,s+1$. Denote the distribution of log censoring times
$\{c_{1},\ldots,c_{n}\}$ by $G$. We define $U_{F_{j}}, U_{H_{j}}$ and $U_{G}$ to be the least upper bound for the support of $F_{j}, H_{j}$ and $G$, respectively.

Throughout the paper the following assumption will be made:
\begin{itemize}
\item[(A1)]  $P(Z_i\leq a_j)= \tau_j$ where $0<\tau_1<\tau_2<\ldots<\tau_s<1$, $E(\ep_i|\mathbf X_i)=0$, $E(T_i^2)<\infty$, $E(\mathbf X_i \mathbf X_i^{\top}  )=\Sigma_0$ is finite and nonsingular, and $Z_i$ and $\ep_i$ are independent, $i=1,\ldots, n$.
\item[(A2)] $T_i$ and $C_i$ are independent, $H_j$ is continuous, and censoring mechanism is independent of covariates.
\item[(A3)] $U_{F_j}<U_{G}$ or $U_{F_j}=U_G=\infty$.
\end{itemize}

Assumption (A1) allows heteroscedastic error. For example, we may relax the error assumption to be $\ep_i=\sigma(\mathbf X_i^{\top}  \mathbf  \beta_j^*)\ep_i^*$ where $\ep_i^*$ is independent with $\mathbf X_i$ and $\ep_1^*,\ldots,\ep_n^*$ are i.i.d. with mean zero and variance $\sigma^2$.  (A2) assumes that $\delta_i$ is conditionally independent of $\mathbf X_i$ given $T_i$.  (A3) implies that $U_{H_j}=U_{F_j}$.

We first present a result for the simple case where $s=1$ is known. Tentatively we rewrite $a_{1}$  and $\mathbf d_1^*$ as $a$ and $\mathbf d^*$, respectively, in model (\ref{cp}), and replace $\tau_1$ with $\tau$ in assumption (A1) for simplicity.

\begin{thm}\label{thm1}
Assume conditions (A1)-(A3) hold for model (\ref{cp}), $s=1$ is known and $\|\mathbf d^*\|>0$.
Let {\small $\hat a=\mbox{argmin}_{\varsigma\in\{Z_{(1)},\ldots,Z_{(n)}\} }\{ \min_{\mathbf \beta}
\frac{\sharp \hat{\mathcal{I}}_{\varsigma^-}}{n}\sum_{i\in \hat{\mathcal{I}}_{\varsigma^-}}w_{[r_{Y_i,\hat{\mathcal{I}}_{\varsigma^-}}]}$ $(Y_{i}-\mathbf X_{i}^{\top}  \mathbf \beta)^2
+ \min_{\mathbf \beta }  \frac{\sharp \hat{\mathcal{I}}_{\varsigma^+}}{n}\sum_{i\in \hat{\mathcal{I}}_{\varsigma^+}}w_{[r_{Y_i,\hat{\mathcal{I}}_{\varsigma^+}}]}(Y_{i}-\mathbf X_{i}^{\top}  \mathbf \beta)^2
\}$}
where  $\hat{\mathcal{I}}_{\varsigma^-}=\{i:Z_i \leq \varsigma\}$ and $\hat{\mathcal{I}}_{\varsigma^+}=\{i:\varsigma<Z_i\}$,
then we have $\frac{1}{n}\sum_{i=1}^n 1_{\{Z_i\leq \hat a\}}\to_{a.s.} \tau$ as $n\to \infty$. Furthermore, if
$Z_i$ is a continuous random variable, we have $\hat a\to_{a.s.} a$; if $Z_i=i$, we have $\hat a/n\to_{a.s.} \tau$.
\end{thm}
The proof of Theorem \ref{thm1} is given in the Appendix. The result itself may be of interest since many econometric studies deal with a single change point.

If $s\geq 0$ is unknown, we use the proposed TSMCD to estimate $s$,  and the change points $a_{j}, j=1, \ldots,s$ when $s>0$. The following technical conditions are also needed:
\begin{itemize}
\item[(A4)]  { $m\to \infty$ and $m=O((n^*)^r)$, where
$0<r\leq 1/2$ is a constant}.
\item[(A5)] The penalty function $p_{\lambda_n,\gamma_n}( |u|)$ of (\ref{esti}) satisfies $p_{\lambda_n,\gamma_n}(0)=0$, $ p_{\lambda_n,\gamma_n}' (u)$ $=0$ if
$u>\gamma_n\lambda_n$, $p_{\lambda_n,\gamma_n}'  (0)=\lambda_n$, $ \lambda_n\rightarrow 0$ and $\lambda_n\sqrt n/\log n\rightarrow\infty$ as $n\to \infty$.
\item[(A6)] $\ep_i=\sigma(\mathbf X_i^{\top}  \mathbf \beta_j^*) \ep_i^*, i\in \mathcal{I}^*_j, j=1,\ldots,s+1$, $\ep_i^*$ is independent of $\mathbf  X_i$, $\ep_1^*,\ldots,\ep_n^*$ are i.i.d. random errors with distribution $F_0$. Suppose
 $U_{F_0}<U_{F_j}$, and  there exists some positive constant $U_{F_j}^*<U_{F_j}-U_{F_0}$,  such that $\sigma(\mathbf X_i^{\top}  \mathbf \beta_j^*)=0$ if $\mathbf X^{\top}   \mathbf \beta_j^*>U_{F_j}^*$, otherwise $\sigma(\mathbf X_i^{\top}  \mathbf \beta_j^*)=1$, $j=1,\ldots,s+1$.
\end{itemize}

{  By law of large numbers and Assumption (A1), $P(Z_i\leq a_j)= \tau_j$ implies $$\sum_{i=1}^n1_{\{a_{j-1}<Z_i\leq  { a}_{j}\}}/n \to_{a.s.} \tau_j-\tau_{j-1}>0.$$  Assumption (A4) implies with probability 1 $$\sum_{i=1}^n1_{\{ \tilde Z_{(n^*-(q_n-j+2)m)}<Z_i\leq \tilde Z_{(n^*-(q_n-j+1)m)}\}}/n = m/n^* \to 0.$$ } {  Thus Assumption  (A1) and (A4)} ensure that
there is at most one threshold in each segment {  $\{Z_i: \tilde Z_{(n^*-(q_n-j+2)m)}<Z_i\leq \tilde Z_{(n^*-(q_n-j+1)m)}\}$} for sufficiently large $n$ where {  $\tilde Z_{(n^*-(q_n-j+2)m)}$ and   $\tilde Z_{(n^*-(q_n-j+1)m)}$, $j=1,\ldots, q_n+1$, are defined in Section 2.1}. Both SCAD and MCD satisfy the penalty assumption (A5). (A6) implies $T_i=\mathbf  X_i^{\top}   \mathbf \beta_j^*$ if $U_{F_j}^*+U_{F_0}<Y_i\leq U_{F_j}$.

Define ${\ma}^*=\{i_1,\ldots,i_a\}$ to be the group index set and $\tilde \bX_{\ma^*}=(\tilde \bX^{(i_1)},\ldots,$ $\tilde \bX^{(i_a)})$
. Similarly we can define
$\hat{\mathbf \theta}_{\ma^*}$ and $\hat{\mathbf \theta}_{(\ma^*)^c}$.
Let $ \ma=\{1,k_{j,n},k_{j,n}+1:j=1,\ldots,s\}$.

\begin{thm} \label{thm2} If Assumptions (A1)-(A6) hold, $\min\{||\mathbf \beta_1^*||,||\mathbf d_1^*||,\ldots,$ $||\mathbf d_s^*||\}$ $>2\sqrt{p} \gamma\lambda_n$, then with probability 1,  (\ref{esti}) has a local minimizer $\hat{\mathbf \theta}$ such that
$\hat{\mathbf \theta}_{\ma^*}=(\tilde \bX_{\ma^*}^{\top}  \tilde \bX_{\ma^*})^{-1}\tilde \bX_{\ma^*}^{\top}\tilde
{\mathbf y}$ and $\hat{\mathbf \theta}_{(\ma^*)^c}=\mathbf 0$,
where
\begin{align}\label{cona}
\ma^*\subseteq \ma~\textrm{and}~1\in \ma^*, k_{\ell,n}\in \ma^*~\textrm{or}~ k_{\ell,n}+1\in \ma^* ~\textrm{for all} ~\ell=1,\ldots,s.
\end{align}
\end{thm}
The proof of Theorem \ref{thm2} is given in Appendix. Theorem \ref{thm2} provides the existence of the solution of (\ref{esti}). In this theorem, $\ma^*$ is the group index set of no-zero group elements of $\hat{\mathbf \theta}$. By (\ref{cona}), $\ma^*$ is not unique and may take $2^s$ possible forms.

{Let $\mathbf {a}=( a_{1},\ldots, a_{s})^{\top}  $ {  be the vector of threshold locations} and $\mathbf {\hat a}=(\hat a_1,\ldots,\hat a_{\hat s})^{\top}$ is obtained by (\ref{esta}). Similar to the definition of $\tilde {\mathbf y}$ and $\tilde \bX$, replacing $\mathcal{I}_j$ with $\mathcal{I}_j^*$ or $\mathcal{\hat I}_j^*$  and then we may define $\tilde {\mathbf y}_{\mathbf {a}}$ or $\tilde {\mathbf y}_{\mathbf {\hat a}}$ and $\tilde \bX_{\mathbf {a}}$ or $\tilde \bX_{\mathbf {\hat a}}$ . By Assumption (A1), we have
$\frac{1}{n}\tilde {\bX}_{\mathbf {a}}^{\top}  \tilde \bX_{\mathbf {a}}\rightarrow_{a.s}\Gamma$, where $\Gamma$ is a positive definite matrix.}

To obtain the limiting property of $\hat {\mathbf \theta}^*$ in (\ref{estit}), we  will further need
the following assumption:
\begin{itemize}
\item[(A7)]  $\max_{u\geq 0}\{p_{\lambda_n,\gamma_n}''  (u)\}+\lambda_{(s+1)p}(\Gamma)>0$ where $\lambda_{(s+1)p}(\Gamma)$ is the minimal eigenvalue of $\Gamma$.
\end{itemize}

For MCP, (A7) is equivalent to $\lambda_{(s+1)p}(\Gamma)>1/\gamma$, and  for SCAD, (A7) is equivalent to $\lambda_{(s+1)p}(\Gamma)>1/(\gamma-1)$.
 Let $ (\theta_1^*,\ldots,\theta_{(s+1)p}^*)^{\top}=$ $
((\mathbf \beta_1^*)^{\top}  ,(\mathbf d_1^*)^{\top} $ ,$ \ldots,(\mathbf d_s^*)^{\top}  )^{\top}$ and $\mathcal{S}=\{j: \theta^*_j\neq 0,j=1,\ldots,(s+1)p\}$.

{  Let $\hat \ma=\{j:\hat {\mathbf \theta}_j\neq 0, j=1,\ldots,q_n+1\}$, and  $\hat \ma^*$ be a subset of $\hat \ma$ such that
$\hat \ma^*=\{j:j\in \hat \ma,j-1\not \in \hat \ma,j=2,\ldots,q_n+1\}=\{\hat k_{1},\ldots, \hat k_{\hat s}\}$  where $\hat s$ is the size of $\hat \ma^*$. }

\begin{thm}\label{thm3}
Under the Assumptions  (A1)-(A7),  $\min_{j\in S}\{|\theta_j|\}>\gamma\lambda_n$,  {  $\hat{\mathbf \theta}$ is the local minimizer in Theorem \ref{thm2}},  {  $Z_i$ is a continuous random variable} and $\hat{\mathbf \theta}^*$ is {  a minimizer of $M(\theta)$} given by (\ref{estit}),
then we have, with probability 1,
\begin{itemize}
\item[(1)]{  $\hat s =s$;}
\item[(2)]{   $\mathbf {\hat a}\to \mathbf {a};$}
\item[(3)] $\hat{\mathbf \theta}^*=\hat{\mathbf \theta}^o$,
\end{itemize}
where $\hat{\mathbf \theta}^o=\arg\min_{\mathbf \theta}\{
\| \tilde {\mathbf y}_{\mathbf {\hat a}}- \tilde \bX_{\mathbf {\hat a}} \mathbf \theta \|^2:\mathbf \theta=(\theta_1,\ldots,\theta_{(s+1)p})^{\top}  ,\theta_j=0 ~\forall~ j\not\in \mathcal S
\}$ is the oracle estimator when thresholds and the set $\mathcal S$ are known.
\end{thm}
The proof of Theorem \ref{thm3} is also given in Appendix. We attain the consistency of the TSMCD estimators in this theorem. Our estimators work as well as the oracle estimators in large samples.

\section{Simulation Study}
We generate random samples from model (\ref{cp}) with $s=2$ thresholds in the following. Specifically, we generate the regressors
$\mathbf X_i=(x_{1,i},x_{2,i},\ldots,x_{6,i})^{\top}$ with $x_{1,i}=1$, $x_{j,i}\sim N(0,1),j=2,\ldots,6$, and $\ep_i\sim N(0,0.5)$. We specify the  coefficients $ (\theta_1^*,\ldots,\theta_{18}^*)^{\top}  =
((\mathbf \beta_1^*)^{\top}  ,(\mathbf d_1^*)^{\top}  ,(\mathbf d_2^*)^{\top}  )^{\top}  =(2,1,1,1,1,1,-1,0,0,-1,-1,-1,0,-1, 1, 0 ,0,0)^{\top}  $, the threshold variable
$Z_i=x_{2,i}$, and true thresholds $a_1=-0.5244$, $a_2=0.2533$. The two thresholds are the 30\% and 60\% lower percentiles of the standard normal distribution. We first design the following four cases:\\
Example 1: $n=150$, $x_{j,i},j=2,\ldots,6$ are independent  and censoring variable $C_i\sim N(2,16)$;\\
Example 2: The same setting as Example 1 except $n=300$;\\
{Example 3: The same setting as Example 2 except $C_i\sim N(\sum_{j=2}^6 x_{j,i},16)$;\\
Example 4: The same setting as Example 3 except covariance between $x_{j_1,i}$ and $x_{j_2,i}$ is $0.5^{|j_1-j_2|}$.}

{The censoring rates are about 40\% for all cases. Assumption (A2) stipulates that $T_i$ and $C_i$ must be independent. This assumption is satisfied in the first two examples. However, a more realistic assumption is that $T_i$ and $C_i$  are conditionally independent given $\mathbf X_i$. Example 3 is thus introduced to test the robustness of our method.} {Example 4 is the case where the regressors are moderately correlated. }

{In order to examine the performance of TSMCP when the model has no threshold. We design the following two cases:\\
  Example 5: The same as Example 1 except  $s=0$ and  $\mathbf \beta_1^*=(1,0,2,0,0,0)$;\\
  Example 6: The same as Example 4 except  $s=0$ and  $\mathbf \beta_1^*=(1,0,2,0,0,0)$;.}

{  All numerical studies are performed on a computer(Intel(R) Core(TM) i7 930 2.80GHz 8 M Caches, 8GB Memory). The {\tt R} program can be downloaded from the first author's website: \url{http://www.stat.nus.edu.sg/~stalj/}.
The means and standard deviations of computing time by TSMCP are reported in Table 1 based on 100 simulations. The average computing time for MCP is  slightly less than that for SCAD. With the increase of model complexity and sample size, the average computing time also tends to increase.
\begin{table}[!htbp]
 \noindent \caption{The means and standard deviations of computation time (seconds). }
\begin{center} \begin{tabular}{cc|cccc|cc}
   \hline
    & &\multicolumn{4}{c|}{$s=2$}&\multicolumn{2}{c}{$s=0$}\\
    \hline
& Example& 1& 2& 3& 4 & 5 &6 \\
\hline
\multirow{2}{1.5cm}{TSMCP-MCP}&Mean&7.55&19.85&21.90&29.56&2.77&7.59\\
&Stand deviation& 1.97&3.15&3.56&5.14&1.59&3.34\\
 \hline
\multirow{2}{1.5cm}{TSMCP-SCAD}&Mean& 9.76&22.58&24.51&30.81&3.61&9.82\\
&Stand deviation& 2.05&3.59&3.50&5.66&1.99&4.04\\
\hline
 \end{tabular}
 \end{center}
 \end{table}
 }

{ The estimation results for $s$ are reported in Table 2 based on 1000 simulations}. Our methods can correctly identify the number of thresholds with very high probability. While their performance is comparable, SCAD seems to identify slightly more unnecessary change points than MCP. Both methods improve as sample size increases.

Figure 2 displays the histograms of the estimators of thresholds, indicating the empirical estimates are symmetrically distributed around the true change points. Table 3 summarizes the estimation performance of the estimated thresholds for the cases with correct estimation of $\hat{s}=s$. Again the two penalty methods give quite similar results. The estimation bias and mean squared error for $a_1$ are relatively smaller than those for $a_2$. In fact we note that the jumps at the two change points are $\|\mathbf d_1^*\|^2=4$ and $\|\mathbf d_2^*\|^2=2$, respectively, under our model. In general it is easier for our methods to detect a break point with greater jump.

\begin{table}[!htbp]
 \noindent \caption{Frequency of estimated $\hat s$ in 1000 simulations. }
\begin{center} \begin{tabular}{cc|ccccccc|cccccc}
   \hline
 & &\multicolumn{7}{c|}{TSMCP-MCP}&\multicolumn{6}{c}{TSMCP-SCAD}\\
    \hline
&$\hat s$ &0   &1  &  2 & 3  & 4& 5 &6 &0  &1   &2   &3  &4 &5\\
\hline
\multirow{4}{0.8cm}{$s=2$}&Example 1&2  &150& 778& 65 & 4& 1 &0  &0  &153 &773 & 70&4 &0\\
&Example 2& 0 &13 & 919& 64 & 3&0  &1  &0  &12  & 869&116&3 &0\\
&Example 3& 0 & 8 &937 & 52 & 3&0  &0  &0  &8   &916 & 71&4 &1\\
&Example 4& 0 & 24&947 &26 & 3&0  &0   &0 &33   &920 & 43&4 &0\\
 \hline
\multirow{2}{0.8cm}{$s=0$}&Example 5& 914 & 82 &4 & 0 & 0&0  &0  &899  &88   &11 & 1&1 &0\\
&Example 6& 945 & 49 &6 & 0 & 0&0  &0  &936  &58   &6 & 0&0 &0\\
\hline
 \end{tabular}
 \end{center}
 \end{table}

 \begin{table}[!htbp]
 \noindent \caption{Estimation performance for the threshold estimation. Bias is the average of estimated parameter minus the true value. RMSE refers to the relative mean squared errors.}
\begin{center} \begin{tabular}{lcccccccc}
   \hline
    &\multicolumn{4}{c}{TSMCP-MCP}&\multicolumn{4}{c}{TSMCP-SCAD}\\
    \hline
    &\multicolumn{2}{c}{$\hat a_{1}$}&\multicolumn{2}{c}{$\hat a_{2}$}
    &\multicolumn{2}{c}{$\hat a_{1}$}&\multicolumn{2}{c}{$\hat a_{2}$}\\
    \hline
    &Bias    &RMSE &Bias&RMSE&Bias&RMSE&Bias&RMSE\\
    \hline
Example 1&  -0.014&0.031&0.037&0.476 &-0.013   &0.037 &0.039&0.428\\
Example 2& -0.007 &0.007&0.010&0.171 &-0.008   & 0.007&0.012&0.158\\
Example 3& -0.009 &0.006&0.008&0.145 &-0.007   &0.006 &0.010&0.159\\
Example 4& -0.006 &0.005&0.013&0.166 &-0.006   &0.005 &0.012&0.158\\
\hline
 \end{tabular}
 \end{center}
 \end{table}

 %\newgeometry{left=3.3cm,right=3.3cm,top=1cm}
\begin{figure}[htb!]\label{fig1}
\begin{center}
{
\centerline
{
\includegraphics[scale=0.45]{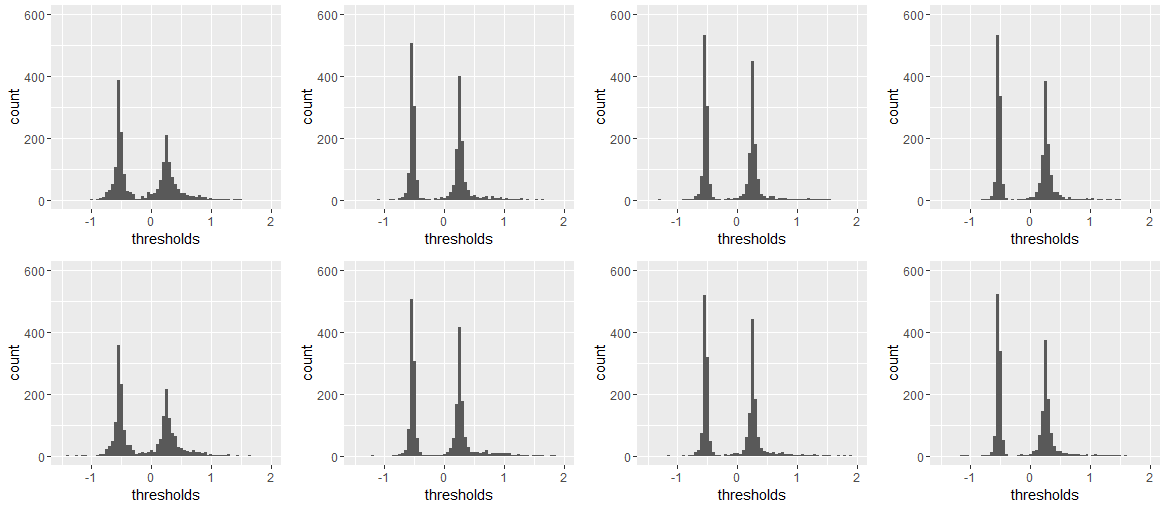}
}
}
\caption{Histograms of the estimated thresholds. The results for MCP and SCAD are displayed in the first and the second row, respectively. Examples 1, 2, 3 and 4  are from the left to the right, respectively.}
\end{center}
\end{figure}

%\restoregeometry

Finally we report the estimation performance of the regression coefficients $\hat\theta_j$ using box-plots in
Figure 3 when the number of thresholds are correctly estimated. In Examples 2-6, the estimated coefficients are quite consistent to the true parameter values. We can see that the variances of $\hat \theta^*_j,j=1,2,7,8,13,14$, which are the estimators of the coefficients of the intercept $x_{1,i}$ and $x_{2,i}=Z_i$, are larger than the others for Examples 1-4. {The zero coefficients $\theta_j^*,j=8,9,13,16,17,18$ in Examples 1-4 and $\theta_j^*,j=2,4,5,6$ in Examples 5-6}, can be successfully identified by our method, suggesting a satisfactory variable selection performance.

{It is noted in all simulations that results from Examples 3-4 and Example 6 are comparable to those from Example 2 and Example 5, respectively.  Even though Examples 3-4 and  Example 6 violate the independence censoring assumption, numerical results in this section suggest our methods are quite robust and may still work steadily under conditional independence.}

\begin{figure}[htb!]\label{fig2}
\begin{center}
{\tiny
\centerline{
\includegraphics[scale=0.45]{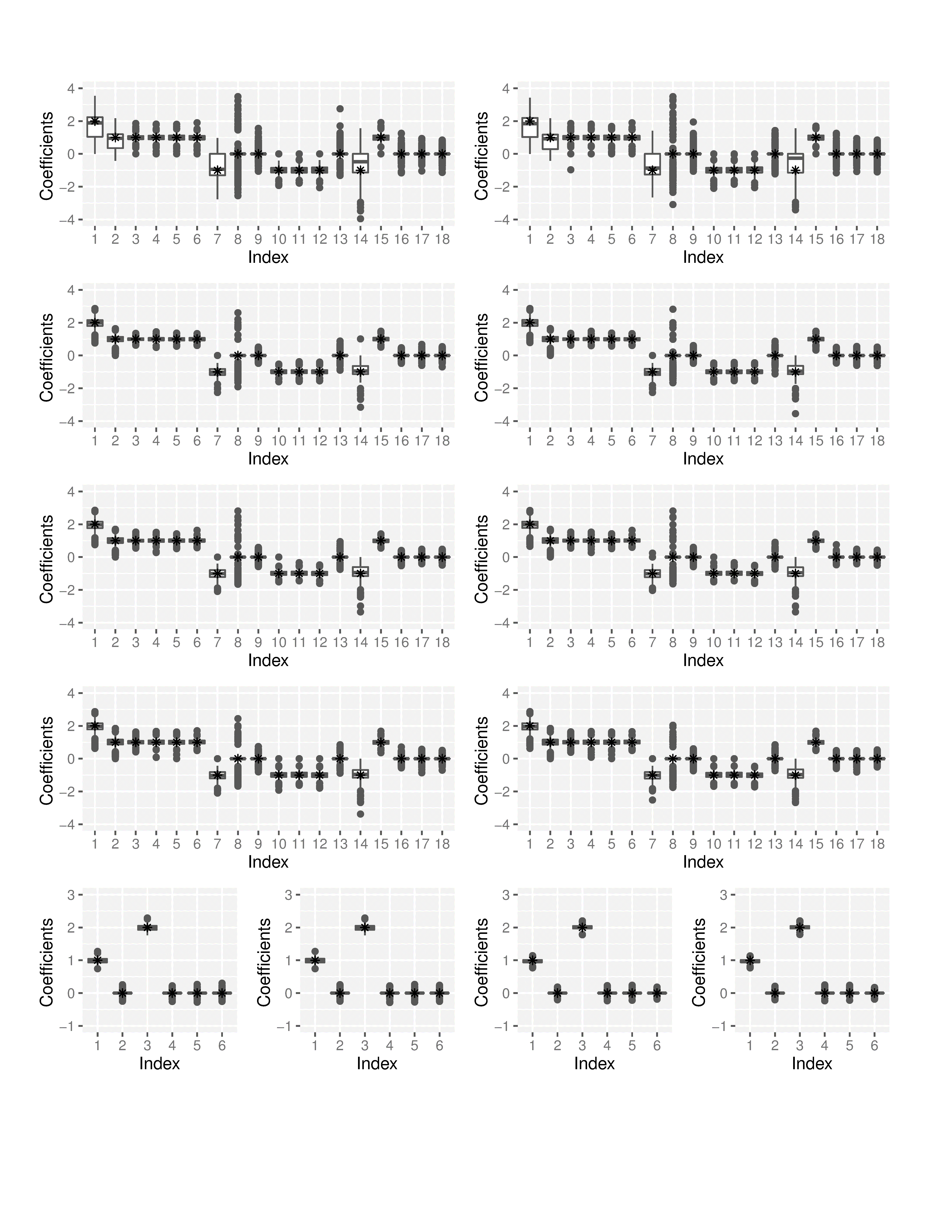}
}
}
\caption{{Box plots of coefficients estimated from MCP (left) and SCAD (right).
Results of examples 1, 2, 3 and 4 are displayed in the four rows, respectively.
Box plots of coefficients estimated from MCP and SCAD of examples 5 and 6 are reported in the five row,
 respectively.
 $" \ast "  $s are the true values.}
}
\end{center}
\end{figure}

\section{Real data analysis}
Follicular lymphoma is the second most common
form of non-Hodgkins lymphoma, accounting for
about 22\% of all documented cases. Dave et al. (2004) investigated whether the survival risks of patients
with follicular lymphoma can be predicted by the
gene-expression profiles of tumors and standard clinical
risk factors at diagnosis. Fresh-frozen
tumor-biopsy specimens and clinical data from 191
untreated patients who had received a diagnosis of
follicular lymphoma between 1974 and 2001 were
obtained. The median age at diagnosis was 51 years
(range 23-81), and the median follow-up time was
6.6 years (range: less than 1.0-28.2). The median
follow-up time among patients alive at last follow-up
was 8.1 years.

A total of 156 subjects are included in
analysis after excluding cases with missing information. Many authors analyzed this data in earlier works. We re-visit this data set in this section and consider an AFT model between the failure time and 5 most significant genetic markers selected in Yu et al. (2012). According to Yu et al. (2012), genes 357, 2345, 6267, 6271, and 3653 in the original sample are the most important markers for the survival risk prediction when clinical information is adjusted. We thus set $x_{1,i}=1$ as the intercept, chose the five gene expressions as the  regressors $x_{2,i},\ldots,x_{6,i}$. {  We pick gene 357 as the index variable in the following analysis and set $Z_i=x_{2,i}$ since it is recognized as the most predictive marker for the failure time. We have also attempted to detect change points for other covariates but our numerical program returned an estimate $\hat{s}=0$ for them.}

  Applying TSMCD{\scriptsize{(MCP)}} and TSMCD{\scriptsize{(SCAD)}},  we obtain the same estimation results, yielding $\hat s=2$ and two thresholds $\hat a_1=-0.483$ and $\hat a_2=0.907$. These two change points divide the sample into three groups with cumulative group sizes $\hat\tau_1=\sum_{i=1}^{156}1_{\{Z_i\leq -0.483\}}/156=0.263$, and $\hat\tau_2=\sum_{i=1}^{156}1_{\{Z_i\leq 0.907\}}/156=0.814$. Figure 4 displays the Kaplan-Meier survival curves for the three groups separated by genes 357. The survival curves for Group 3 drops rapidly since the baseline and represents a high-risk group in this sample.
 Such an observation is not available without considering the threshold AFT model in this paper. Practitioners may adapt our model easily to discover more meaningful sub-populations with defining features.

{  Traditionally when the goal is to identify subgroups from the empirical sample one may resort  to clustering analysis (Gordon (1981)). Such unsupervised  learning methods produce segmentation or grouping based on the variation of covariates and ignore the censored response variable. One drawback is that those learned groupings may not be relevant to the survival response. We have implemented a hierarchical clustering procedure ({\tt R} function {\tt hclust}) using only the covariate matrix and the identified clusters are quite different from our change point analysis. The comparison of cluster sizes between clustering analysis and our method is given in the following table.
\begin{table}[!htbp]
 \noindent \caption{Cluster sizes of real data analysis.  }
\begin{center} \begin{tabular}{c|ccc}
   \hline
  &\multicolumn{3}{c}{Cluster analysis}\\
 {TSMCD} & Group 1 & Group 2 & Group 3\\
Group 1 & 32 & 5 & 4\\
Group 2 & 20 & 46 & 20 \\
Group 3 & 0 & 14 & 15\\
\hline
 \end{tabular}
 \end{center}
 \end{table}
There is only slight agreement between the two methods with Cohen's kappa value equal to $0.36$.  Furthermore, it is hard to interpret the groups resulted from the hierarchical clustering algorithm. In particular it is unclear if the three groups suggest any practical difference in survival probability. Examining the Kaplan-Meier curves for the three groups are shown in Figure 5, we notice that the survival functions for groups 2 and 3 severely overlap for the observation period. Such findings might be less meaningful for clinical investigation.}

\begin{figure}[htb!]\label{fig3}
\begin{center}
{\tiny
\centerline{
\includegraphics[scale=0.35]{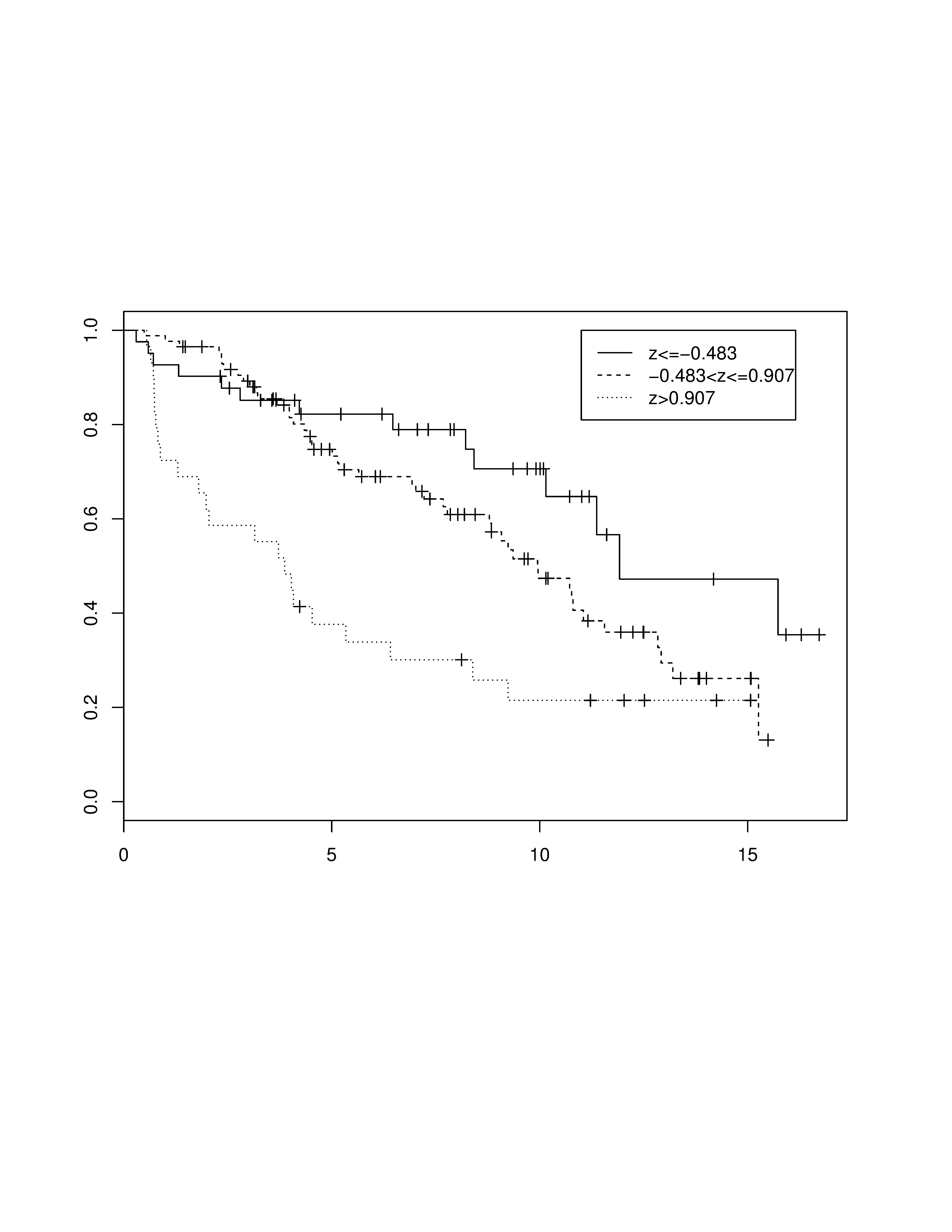}
}}
\caption{Survival curves for the three groups separated by the two thresholds of Gene 357. $+$ indicates a censored observation. Groups 1, 2 and 3 refer to $Z\leq -0.482 $, $-0.482<Z\leq 0.907 $ and $Z>0.907$, respectively.}
\end{center}
\end{figure}

The estimated regression coefficients are $\hat {\mathbf \theta}^*=((\hat {\mathbf \beta}_1^*)^{\top}  , (\hat {\mathbf d}_1^*)^{\top}  , (\hat {\mathbf d}_2^*)^{\top}  )^{\top}   $ where $\hat {\mathbf \beta}_1^*=(0.791, -0.492$,  $0.291$, $0.990$, $-0.589,  1.115)^{\top}  $, $\hat {\mathbf d}_1^*=(1.166, 0, -0.451$, $-1.055,  0.540, -0.991)^{\top}  $, $\hat {\mathbf d}_2^*=( 0$, $-0.381$, $0$,  $0$,  $0$,  $0.770)^{\top}  $. Such results suggest at the first threshold $\hat{a}_1=-0.483$, only the coefficient of $x_{2,i}$ does not change; and at the second threshold $\hat{a}_2=0.907$, only coefficients of $x_{2,i}$ and $x_{6,j}$ change and all other coefficients remain the same.

 Using the identified break point structure, we refitted the AFT model and reported the estimation results for the three sub-groups in Table 4, along with 95\% bootstrap confidence intervals (Huang et al. (2006)). The effects of genes on the survival outcome are quite different for the three groups. For example, Gene 2345 is not significant for the first group but are negatively associated with the failure time for the other two groups. Using our empirical findings, investigators may evaluate the effects of various genetic biomarkers on mortality more specifically for different sub-populations.

 \begin{table}[!htbp]
 \noindent \caption{Estimates of regression coefficients for the three groups, along with bootstrap standard errors (S.E.) and P-values by Wald test.}
\begin{center} \begin{tabular}{lccccccccc}
   \hline
   &\multicolumn{3}{c}{$Z\leq -0.482 $}&\multicolumn{3}{c}{$-0.482<Z\leq 0.907 $}&\multicolumn{3}{c}{$Z>0.907$}\\
    \hline
Covariates    &Coef.  &S.E.& P-value  & Coef.  &S.E. & P-value  & Coef.  &S.E.& P-value \\
\hline
Intercept& 0.791  &0.507& 0.056    &1.957 & 0.083& 0     & 1.957  &0.083& 0     \\
Gene 357      & -0.492 &0.179& 0.003    &-0.492& 0.179& 0.003     & -0.874 &0.144& 6e-10      \\
Gene 2345     & 0.291  &0.627&0.321 &-0.159&0.084 &0.029      & -0.159 &0.084&0.029  \\
Gene 6267     & 0.990  &0.725&0.086     &-0.065& 0.110&0.277     & -0.065 &0.110&0.277      \\
Gene 6271     & -0.589 &0.568&0.150&-0.049& 0.120&0.341&  -0.049&0.120&0.341 \\
Gene 3653     & 1.115  &0.556&0.022     &0.124 & 0.078&0.056      &  0.894 &0.250&2e-4      \\
\hline
 \end{tabular}
 \end{center}
 \end{table}

\begin{figure}[htb!]\label{fig5}
\begin{center}
{\tiny
\centerline{
\includegraphics[scale=0.35]{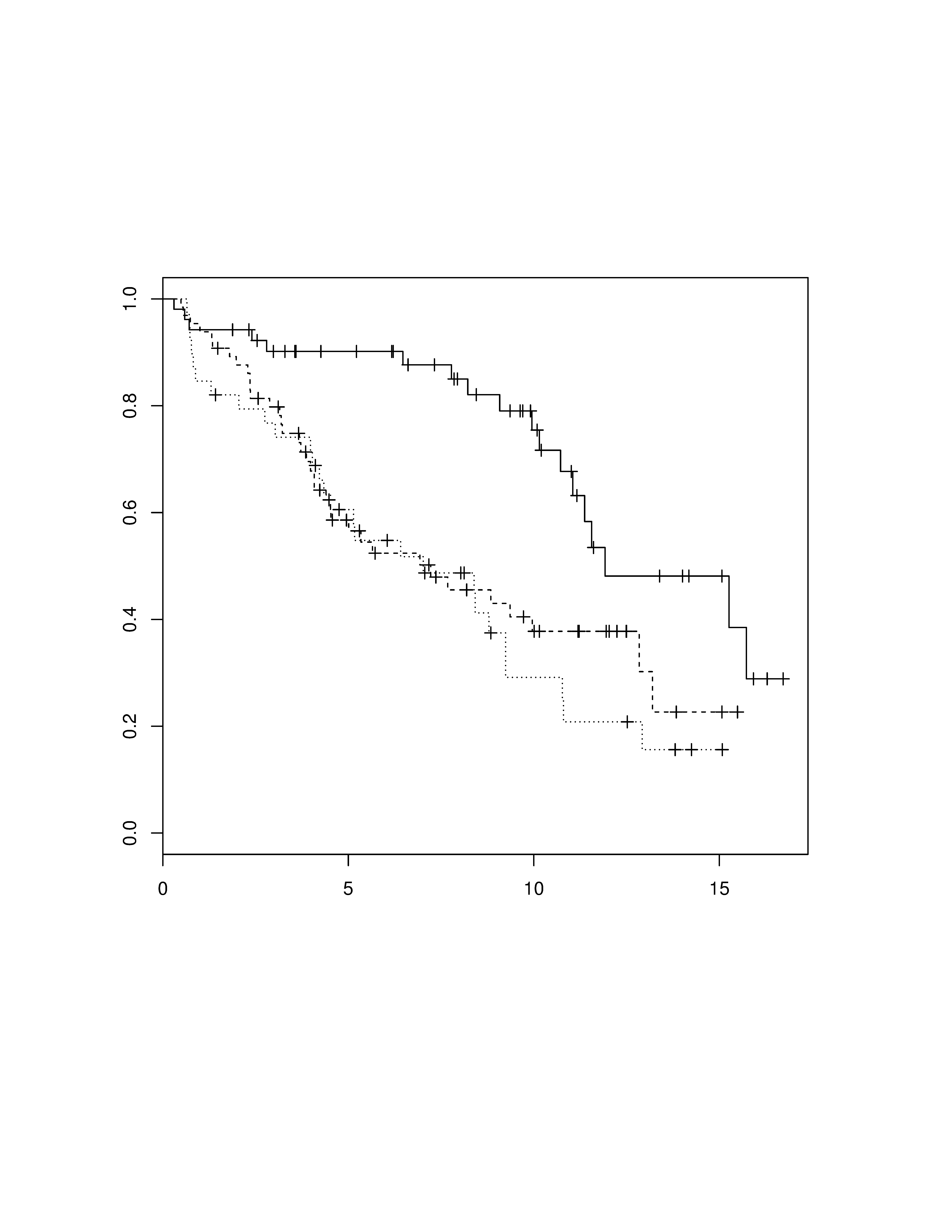}
}}
\caption{Kaplan-Meier curves from the hierarchical clustering algorithm.}
\end{center}
\end{figure}

\section{Discussion}
{
Stute estimator for AFT model is mainly developed for right censored survival data. When the failure time is interval censored, usually a likelihood-based objective function is constructed to facilitate the regression estimation. One can incorporate change point structure in such framework and carry out the two-step estimation similar to our proposal. However, we note that the functional form for the likelihood under interval censoring may be quite complicated and the numerical solution may not be straightforward. More extensions in theory and computation are needed to provide a complete solution.

We consider independent survival times in this paper. Individuals observed over time may experience multiple events of the same type, or a number of events of different types. Correlated survival time data require further model structures and technical assumptions to achieve efficient results. Two familiar approaches are the marginal model and the frailty model (Kalbfleisch and Prentice (2002), ch 10). In both approaches one may still consider Stute estimator for the AFT regression and our TSMCD adapts directly. However, the asymptotic results for such analysis are not trivial since one must take care of the inter-dependence of the repeated measures.

The TSMCD involves the penalization approaches and thus can be easily extended to incorporate high-dimensional data analysis. Not many authors examined the change point problem under such settings and therefore the results in this paper could contribute significantly towards this goal. However, when the dimension of feature space is ultra high with an exponential order of the sample size, usually a screening step must be conducted before the application of penalized estimation. In particular, one may adopt TSMCD at the screening step and estimate the effects of individual markers using the change point structure. This differs from the existing literature where the marginal effects are usually modeled as a smooth function at the initial screening. The detailed methodology construction is beyond the scope of this paper. We will carry out necessary theoretical and empirical studies for this topic.}

\appendix

\section{Matrix representation  of model (2.1)}\label{app1}

Let $\mathcal{C}_{i}=\bigcup_{j=i}^{q_n+1}\mathcal{I}_{j}$, $i=1,\ldots,q_n+1$.
We split the segment $\mathcal{I}_{k_{j,n}}$ into two segments: $\mathcal{I}_{k_{j,n}}^{(1)}=\{i: \tilde Z_{(n^*-(q_n-k_{j,n}+2)m))}<Z_i\leq a_j\}$
and $\mathcal{I}_{k_{j,n}}^{(2)}=\{i:a_j<Z_i\leq \tilde Z_{(n^*-(q_n-k_{j,n}+1)m))}\}$.
We rewrite the model (\ref{cp}) as the following
linear regression model that takes the segmentation of the data sequence into consideration:
\begin{align}\label{newcp1}
T_i=\mathbf X_i^{\top}  \left(\mathbf \beta_1^*+\sum_{j=1}^{s}\mathbf d_{j}^*
1_{\{i\in\mathcal{C}_{k_{j,n}}\}}
\right)-\mathbf X_i^{\top}  \sum_{j=1}^{s}\mathbf d_{j}^*
1_{\{i\in\mathcal{I}_{k_{j,n}}^{(1)}\}}+\varepsilon_i,
\end{align}
where  $i=1,\ldots,
n$, $\mathbf d_{j}^*=\mathbf \beta_{j+1}^*-\mathbf \beta_j^*$, $j=1,\ldots,s$.
 To include them in (\ref{newcp1}) is to facilitate the proof
of Theorem \ref{thm2}.
In order to estimate $k_{j,n}$, $j=1,\ldots,s$ and the regression coefficients simultaneously,  we expand the model (\ref{newcp1}) as the following model:
\begin{equation}
T_i=\mathbf X_i^{\top}  \left[\mathbf \beta_1+\sum_{\ell=1}^{q_n}\mathbf d_{\ell}
1_{\{i\in\mathcal{C}_{\ell+1}\}}+\mathbf \omega_i
\right]+\varepsilon_i,\quad i=1,\ldots,
n,\label{newcp}
\end{equation}
where $\{\mathbf d_\ell\}$ and
$\{\mathbf \omega_i\}$ are defined as follows:
\begin{align*}
\mathbf d_\ell&=\left\{\begin{array}{ll}
\mathbf d_j^*,& \ell=k_{j,n},\ j=1,\ldots,s\\
\mathbf 0_p,&  \textrm{otherwise} ;\end{array}\right.\\
\mathbf \omega_i&=\left\{\begin{array}{ll}
-\mathbf d_j^*,& i\in \mathcal{I}_{k_{j,n}}^{(1)},\ j=1,\ldots,s\\
\mathbf 0_p,&  \textrm{otherwise} .\end{array}\right.\end{align*}

Let $\mathbf \ep_{(j)}=(\ep_{i},i\in \mathcal{I}_j)^{\top}  $, $\mathbf x_{\omega,(j)}=(\mathbf X_i^{\top}  \mathbf \omega_i,i\in \mathcal{I}_j)^{\top}  $,
$\tilde{\mathbf \ep}_{(j)}=diag(\tilde  {\mathbf w}_{(j)})^{1/2}\mathbf \ep_{(j)}$,  $\tilde{\mathbf x}_{\omega,(j)}=diag(\tilde  {\mathbf w}_{(j)})^{1/2}\mathbf x_{\omega,(j)}$, $\tilde {\mathbf \ep}=(\mathbf \ep_{(1)}^{\top}, \ldots, \mathbf \ep_{(q_n+1)}^{\top}  )^{\top}  $,
$\tilde {\mathbf x}_{\omega}=(\tilde{\mathbf x}_{\omega,(1)}^{\top}  ,\ldots,\tilde{\mathbf x}_{\omega,(q_n+1)}^{\top}  )^{\top}  $, $\mathbf \theta^*=(\mathbf \theta_1^{\top}  ,\ldots,
\mathbf \theta_{q_n+1}^{\top}  )^{\top}  =(\mathbf \beta^{\top}  _1,\mathbf d_1^{\top}  ,\ldots,\mathbf d_{{q_n}}^{\top}  )^{\top}  $ and
$\mathbf d_j=(d_{j1}, \ldots,d_{jp})^{\top}  $, $j=1,\ldots,{q_n}$.
The model (\ref{newcp}) has the matrix form as follows:
 \begin{align}\tilde {\mathbf y}=\tilde \bX
{\mathbf \theta}^*+\tilde {\mathbf x}_\omega+\tilde {\mathbf \ep},\label{regy}\end{align}

Remove the zero elements of ${\mathbf \theta}^*$, model (\ref{regy}) can be rewritten as
 \begin{align}\tilde {\mathbf y}=\tilde \bX_{\ma}
{\mathbf \theta}_{\ma}^*+\tilde {\mathbf x}_\omega+\tilde {\mathbf \ep},\label{regy0}\end{align}
which corresponds to the matrix form of model (\ref{newcp1}).

\section{Proof of Theorems 1-3}

\subsection{Proof of Theorem \ref{thm1}}\label{app2}

Let $\varsigma=Z_{(k)}$, $\mathbf Y_{1k}=(Y_i,i\in \mathcal{I}_{Z_{(k)}^{-}})^{\top}  $, $\bX_{1k}=(\mathbf X_i,i\in \mathcal{I}_{Z_{(k)}^{-}})^{\top}  $, $\mathbf  \ep_{1k}=(\ep_i,i\in \mathcal{I}_{Z_{(k)}^{-}})^{\top}  $, $\mathbf Y_{2k}=(Y_i,i\in \mathcal{I}_{Z_{(k)}^{+}})^{\top}  $, $\bX_{2k}=(\mathbf X_i,i\in \mathcal{I}_{Z_{(k)}^{+}})^{\top}  $, $\mathbf \ep_{2k}=(\ep_i,i\in \mathcal{I}_{Z_{(k)}^{+}})^{\top}  $, $\bW_{1k}=diag((w_{[r_{Y_i,\mathcal{I}_{Z_{(k)}^{-}}}]},i\in \mathcal{I}_{Z_{(k)}^{-}}))$, $\bW_{2k}=diag((w_{[r_{Y_i,\mathcal{I}_{Z_{(k)}^{+}}}]},i\in \mathcal{I}_{Z_{(k)}^{+}}))$, $Q_1(k,\mathbf \beta)=Q(Z_{(k)}^{-},\mathbf \beta)$ and $Q_2(k,\mathbf \beta)=Q(Z_{(k)}^{+},\mathbf \beta)$ .
Then we have
\begin{align*}
\hat{\mathbf {\beta}}_{1k}&=\arg \min_{\mathbf \beta}Q_1(k,\mathbf \beta)= (\bX_{1k}^{\top}  \bW_{1k}\bX_{1k})^{-1}\bX_{1k}^{\top}  \bW_{1k}\mathbf  Y_{1k}\\
\hat{\mathbf {\beta}}_{2k}&=\arg \min_{\mathbf \beta}Q_2(k,\mathbf \beta)= (\bX_{2k}^{\top}  \bW_{2k}\bX_{2k})^{-1}\bX_{2k}^{\top}  \bW_{2k}\mathbf Y_{2k}.
\end{align*}

Let $k/n\to \tau'  $, $k^*/n\to \tau$ (for example $k=\lfloor n \tau'   \rfloor$, $k^*=\lfloor n \tau\rfloor$), $E(\mathbf X_i\mathbf X_i^{\top}  |Z_i\leq \varsigma)=\Sigma_{0,\varsigma^{-}}$
and $E(\mathbf X_i\mathbf X_i^{\top}  |Z_i>\varsigma)=\Sigma_{0,\varsigma^{+}}$.
First  we consider the case of $ 0\leq \tau'  <\tau$. For $\sharp\mathcal{I}_{Z_{(k)}^{-}}=k$,  we get
\begin{align*}
Q_1(k,\hat{\mathbf {\beta}}_{1k})
&=\frac{k}{n}(\mathbf Y_{1k}-\bX_{1k} \hat{\mathbf {\beta}}_{1k})^{\top}  \bW_{1k}(\mathbf Y_{1k}-\bX_{1k} \hat{\mathbf {\beta}}_{1k})\\
&=\frac{k}{n}\mathbf Y_{1k}^{\top}  \left(I_{k}-\bW_{1k}\bX_{1k} (\bX_{1k}^{\top}  \bW_{1k} \bX_{1k})^{-1}\bX_{1k}^{\top}  \right)\\
&\bW_{1k}\left(I_{k}-\bX_{1k} (\bX_{1k}^{\top}   \bW_{1k} \bX_{1k})^{-1}\bX_{1k}\bW_{1k}\right)\mathbf Y_{1k}\\
&=\frac{k}{n}\mathbf \ep_{1k}^{\top}   \bW_{1k}^{1/2}(I_{k}-\bW_{1k}^{1/2}\bX_{1k} (\bX_{1k}^{\top}  \bW_{1k} \bX_{1k})^{-1}\bX_{1k}^{\top}  \bW_{1k}^{1/2})\bW_{1k}^{1/2}\mathbf \ep_{1k}\\
&=\frac{k}{n}\mathbf \ep_{1k}^{\top}   \bW_{1k}\mathbf \ep_{1k}-\frac{k}{n}\mathbf \ep_{1k}^{\top}  \bW_{1k}\bX_{1k} (\bX_{1k}^{\top}  \bW_{1k} \bX_{1k})^{-1}\bX_{1k}^{\top}  \bW_{1k}\mathbf \ep_{1k}.
\end{align*}
By  Corollary 1.8  of Stute (1993) and the assumption that $Z_i,\ep_i$ are independent,
{\small
\begin{align*}
\mathbf \ep_{1k}^{\top}   \bW_{1k}\mathbf \ep_{1k}&=\sum_{i\in \mathcal{I}_{Z_{(k)}^{-}}} w_{[r_{Y_i,\mathcal{I}_{Z_{(k)}^{-}}}]}(Y_i-\mathbf X_i^{\top}  \mathbf {\beta}_1^*)^2\rightarrow_{a.s.}
E[(Y_i-\mathbf X_i^{\top}  \mathbf {\beta}_1^*)^2|Z_i\leq Z_{(k)}]\\
& =E\ep_i^2=\sigma_1^2,\\
\bX_{1k}^{\top}  \bW_{1k} \bX_{1k}&=\sum_{i\in \mathcal{I}_{Z_{(k)}^{-}}} w_{[r_{Y_i,\mathcal{I}_{Z_{(k)}^{-}}}]}\mathbf X_i\mathbf X_i^{\top}  \rightarrow_{a.s.} E(\mathbf X_i\mathbf X_i^{\top}  |Z_i\leq Z_{(k)})=\Sigma_{0,Z_{(k)}^{-}}\leq \Sigma_0,\\
\mathbf \ep_{1k}^{\top}  \bW_{1k}\bX_{1k}&=\sum_{i\in \mathcal{I}_{Z_{(k)}^{-}}} w_{[r_{Y_i,\mathcal{I}_{Z_{(k)}^{-}}}]}\ep_i\mathbf  X_i^{\top}  \rightarrow_{a.s.} E(\ep_i\mathbf X_i^{\top}  |Z_i\leq Z_{(k)})\\
&=E[E(\ep_i\mathbf X_i^{\top}  |\mathbf X_i,Z_i\leq Z_{(k)})]=\mathbf 0_p^{\top}  .
\end{align*}
}
Thus \begin{align}\label{q1k}
Q_1(k,\hat{\mathbf {\beta}}_{1k})\to_{a.s.} \tau' \sigma_1^2.
\end{align}

Let $\mathbf d^*=\mathbf \beta_2^*-\mathbf \beta_1^*$, $\mathbf T_{2k}=(T_{k+1},\ldots,T_{n})^{\top}  $,$\bbzero_{kk^*}=\bbzero_{(k^*-k)\times (p+1)}$ and $\bX_{2k^*}=(\mathbf X_{k^*+1},\ldots,\mathbf X_n)^{\top}  $. We have \begin{align*}\mathbf T_{2k}=\bX_{2k} \mathbf  \beta_1^*+\begin{pmatrix}\bbzero_{kk^*}\\ \bX_{2k^*}\end{pmatrix} \mathbf d^* +\mathbf \ep_{2k}.\end{align*}

Notice that $\sharp\mathcal{I}_{Z_{(k)}^+}=n-k$, we obtain
\begin{align*}
Q_2(k,\hat{\mathbf {\beta}}_{2k})&=\frac{n-k}{n}(\mathbf Y_{2k}-\bX_{2k} \hat{\mathbf {\beta}}_{2k})^{\top}  \bW_{2k}(\mathbf Y_{2k}-\bX_{2k} \hat{\mathbf {\beta}}_{2k})\\
&=\frac{n-k}{n}\mathbf Y_{2k}^{\top}  \bW_{2k}^{1/2}\bP_{n-k}\bW_{2k}^{1/2}\mathbf Y_{2k}\\
&=\frac{n-k}{n}\mathbf \ep_{2k}^{\top}   \bW_{2k}^{1/2}\bP_{n-k}\bW_{2k}^{1/2}\mathbf \ep_{2k}\\
&+\frac{n-k}{n}(\mathbf d^*)^{\top}  \begin{pmatrix}\bbzero_{kk^*}\\ \bX_{2k^*}\end{pmatrix}^{\top}  \bW_{2k}^{1/2}\bP_{n-k}\bW_{2k}^{1/2}\begin{pmatrix}\bbzero_{kk^*}\\ \bX_{2k^*}\end{pmatrix}\mathbf d^*\\
&+\frac{2(n-k)}{n}(\mathbf d^*)^{\top}  \begin{pmatrix}\bbzero_{kk^*}\\ \bX_{2k^*}\end{pmatrix}^{\top}  \bW_{2k}^{1/2}\bP_{n-k}\bW_{2k}^{1/2}\mathbf \ep_{2k}
\end{align*}
where $\bP_{n-k}=I_{n-k}-\bW_{2k}^{1/2}\bX_{2k} (\bX_{2k}^{\top}  \bW_{2k} \bX_{2k})^{-1}\bX_{2k}^{\top}  \bW_{2k}^{1/2}$.

Recall that $k/n \to \tau' , k^*/n \to \tau, 0\leq \tau'  <\tau$. Let $\alpha_{1k}=\lim_{n\to \infty} (k^*-k)/(n-k)=(\tau-\tau' )/(1-\tau' )$. In the following proof, we can consider that $\{T_{i},i\in \mathcal{I}_{Z_{(k)}^+}\}$ are independent samples with the same mixture distribution as $\tilde F_k=\alpha_{1k} F_1+(1-\alpha_{1k})F_2$.
 Let the distribution of  $Y_i=\min\{T_i,C_i\}$, $i\in \mathcal{I}_{Z_{(k)}^+}$ be $\tilde H_k$.
By (A2),
\begin{align*}
1-\tilde H_k(y)&=(1-\tilde F_k(y))(1-G(y))\\
&=\alpha_{1k} (1-F_1(y))(1-G(y))+(1-\alpha_{1k})(1-F_2(y))(1-G(y))\\ &=\alpha_{1k}(1-H_1(y))+(1-\alpha_{1k})(1-H_2(y)),
\end{align*}
which shows that $Y_i,i\in \mathcal{I}_{z_{(k)}^+}$ are independent samples with the same mixture  distribution $\tilde H_k(y)=\alpha_{1k} H_1(y)+(1-\alpha_{1k}) H_2(y)$.

Let $\tilde {\mathbf {\beta}}^*=\mathbf \beta_1^* 1_{\{T\overset{d}{=}F_1\}}+\mathbf \beta_2^*1_{\{T\overset{d}{=}F_2\}}$.
By Corollary 1.8  of Stute (1993), we obtain
\begin{align*}
\mathbf \ep_{2k}^{\top}   \bW_{2k}\mathbf \ep_{2k}&=\sum_{i\in\mathcal{I}_{Z_{(k)}^+}} w_{[r_{Y_i,\mathcal{I}_{Z_{(k)}^+}}]}(Y_i-\mathbf X_i^{\top}   \tilde {\mathbf {\beta}}^*)^2\\
&\rightarrow_{a.s.}E[(T_i-\mathbf X_i^{\top}   \tilde {\mathbf {\beta}}^*)^2|Z_i>Z_{(k)}]=\alpha_{1k}\sigma_1^2+(1-\alpha_{1k})\sigma_2^2,\\
\bX_{2k}^{\top}  \bW_{2k} \bX_{2k}&\rightarrow_{a.s.} E(\mathbf X_i\mathbf X_i^{\top}  |Z_i>Z_{(k)})=\Sigma_{0,Z_{(k)}^+}\leq \Sigma_0,\\
\mathbf \ep_{2k}^{\top}  \bW_{2k}\bX_{2k}&\rightarrow_{a.s.} E(\ep_i\mathbf X_i^{\top}  |Z_i>Z_{(k)})=\mathbf 0_p^{\top}  .
\end{align*}

By Theorem 1 of Stute (1993), we have
\begin{align*}
\bX_{2k}^{\top}  \bW_{2k}\begin{pmatrix}\bbzero_{kk^*}\\ \bX_{2k^*}\end{pmatrix}\mathbf d^*&=\sum_{i\in\mathcal{I}_{Z_{(k^*)}^+}} w_{[r_{Y_i,\mathcal{I}_{Z_{(k)}^+}}]}\mathbf X_i\mathbf X_i^{\top}   \mathbf d^*\\
&=\sum_{i\in \mathcal{I}_{Z_{(k)}^+}} w_{[r_{Y_i,\mathcal{I}_{Z_{(k)}^+}}]}\mathbf X_i\mathbf X_i^{\top}   1_{\{T_i\overset{d}{=}F_2\}}\mathbf d^*\\
&\rightarrow_{a.s.}E[(\mathbf X_i\mathbf X_i^{\top}  1_{\{T_i\overset{d}{=}F_2\}})|Z_i>Z_{(k^*)}]\mathbf  d^*=(1-\alpha_{1k})\Sigma_{0,Z_{(k^*)}^+}\mathbf  d^*.
\end{align*}
Similarly, we have
{\small\begin{align*}
(\mathbf d^*)^{\top}  \begin{pmatrix}\bbzero_{kk^*}\\ \bX_{2k^*}\end{pmatrix}^{\top}  \bW_{2k}\begin{pmatrix}\bbzero_{kk^*}\\ \bX_{2k^*}\end{pmatrix}\mathbf d^*&=(\mathbf d^*)^{\top}  \sum_{i\in\mathcal{I}_{Z_{(k^*)}^+}} w_{[r_{Y_i,\mathcal{I}_{Z_{(k)}^+}}]}\mathbf X_i\mathbf X_i^{\top}   \mathbf d^*\\
&\rightarrow_{a.s.}(1-\alpha_{1k})(\mathbf d^*)^{\top}  \Sigma_{0,Z_{(k^*)}^+}\mathbf d^*,\\
(\mathbf d^*)^{\top}  \begin{pmatrix}\bbzero_{ka_n}\\ \bX_{2k^*}\end{pmatrix}^{\top}  \bW_{2k}\mathbf \ep_{2k}&=(\mathbf d^*)^{\top}  \sum_{i\in\mathcal{I}_{Z_{(k^*)}^+}} w_{[r_{Y_i,\mathcal{I}_{Z_{(k)}^+}}]}\mathbf X_i \ep_{i}\\
&\rightarrow_{a.s.}(1-\alpha_{1k})(\mathbf d^*)^{\top}  E(\mathbf X_i \ep_{i}|Z_i>Z_{(k^*)})=0.
\end{align*}
}
Thus
{\small
\begin{align}\label{q2k}
 \quad \quad \quad Q_2(k,\hat\beta_{2k})\to_{a.s.}
(\tau-\tau')\sigma_1^2+(1-\tau)\sigma_2^2+\frac{(1-\tau)(\tau-\tau')}{1-\tau'}(\mathbf d^*)^{\top}\Sigma_{0,z_{(k^*)}^+}\mathbf d^*.
\end{align}
}
Combining (\ref{q1k}) and (\ref{q2k}), if $0\leq \tau'  <\tau$, we have, with probability 1, {\small $$Q(Z_{(k)})\to \tau\sigma_1^2+(1-\tau)\sigma_2^2+(1-\tau)(\tau-\tau' )(1-\tau'  )^{-1}(\mathbf d^*)^{\top}  \Sigma_{0,Z_{(k^*)}^+}\mathbf d^*>\tau\sigma_1^2+(1-\tau)\sigma_2^2.$$}

Similarly, if $1\geq \tau' >\tau$, with probability 1,  we can obtain
$$Q(Z_{(k)})\to \tau\sigma_1^2+(1-\tau)\sigma_2^2+\tau' (1-\tau')^{-1}(\mathbf d^*)^{\top}  \Sigma_{0,Z_{(k^*)}^-}\mathbf d^*>\tau\sigma_1^2+(1-\tau)\sigma_2^2,$$ and
 $Q(Z_{(k^*)})\to_{a.s.} \tau\sigma_1^2+(1-\tau)\sigma_2^2$. Thus if $\hat a=\arg \min_{\varsigma\in\{Z_{(1)},\ldots,Z_{(n)}\} }\{Q(\varsigma)\}=Z_{(\hat k)}$, we have $\hat k/n\to_{a.s} \tau$, i.e., $\frac{1}{n}\sum_{i=1}^n1_{\{Z_i\leq \hat a\}}\to_{a.s} \tau$. If $Z_i$ is a continuous random variable, by continuous mapping theorem, we have $\hat a\to_{a.s.} a$. Therefore, the proof is completed.\\

\par

\bigskip

\subsection{Proof of Theorem \ref{thm2}}\label{app3}

By the Karush-Kuhn-Tucker (KKT) conditions, we only need to show with probability 1,
\begin{align}
\max_{j\not\in \ma}||n^{-1}(\tilde \bX^{(j)})^{\top}  (\tilde
{\mathbf y}-\tilde \bX_{\ma^*}\hat{\mathbf \theta}_{\ma^*})||\leq \sqrt{p}\lambda_n \label{KKT1}\\
\min_{\ell=1,\ldots,s}\{\|\hat{\mathbf \theta}_1\|,\max\{\|\hat{\mathbf \theta}_{k_{\ell,n}}\|,\|\hat{\mathbf \theta}_{k_{\ell,n}+1}\|\}\}\geq \sqrt{p}\gamma\lambda_n.\label{KKT2}
\end{align}

First, we prove (\ref{KKT1}). Note that $\alpha_j$ is arbitrary in (\ref{regy0}). By the condition (A7) and (A8), we put $\alpha_j=1$ if $k_{j,n}\in \ma^*$ and $k_{j,n}+1\not\in \ma^*$, and
we put $\alpha_j=0$ if $k_{j,n}\not\in \ma^*$ and $k_{j,n}+1\in \ma^*$. Thus, (\ref{regy0}) can be rewritten as \begin{align*}
\tilde
{\mathbf y}=\tilde \bX_{\ma^*}
\tilde
{\mathbf \theta}_{\ma^*}^*+ \tilde {\mathbf x}_\omega+\tilde {\mathbf \ep}.
\end{align*}

By $\hat{\mathbf \theta}_{\ma^*}=(\tilde \bX_{\ma^*}^{\top}  \tilde \bX_{\ma^*})^{-1}\tilde \bX_{\ma^*}^{\top} \tilde
{\mathbf y}_{\ma^*}$, we obtain
\begin{align*}
&\|(\tilde \bX^{(j)})^{\top}  (\tilde
{\mathbf y}-\tilde \bX_{\ma^*}\hat{\mathbf \theta}_{\ma^*})\|\\
&=\|(\tilde \bX^{(j)})^{\top}  (I_n-\tilde \bX_{\ma^*}(\tilde \bX_{\ma^*}^{\top}  \tilde \bX_{\ma^*})^{-1}\tilde \bX_{\ma^*}^{\top} )\tilde
{\mathbf y}\|\\
&=\|(\tilde \bX^{(j)})^{\top}  (I_n-\tilde \bX_{\ma^*}(\tilde \bX_{\ma^*}^{\top}  \tilde \bX_{\ma^*})^{-1}\tilde \bX_{\ma^*}^{\top} )\tilde
(\tilde {\mathbf x}_\omega+\tilde {\mathbf \ep})\|\\
&\leq\|(\tilde \bX^{(j)})^{\top}  \tilde
(\tilde {\mathbf x}_\omega+\tilde {\mathbf \ep})\|\leq \|(\tilde \bX^{(j)})^{\top}
\tilde {\mathbf x}_\omega\|+\|(\tilde \bX^{(j)})^{\top}  \tilde {\mathbf \ep}\|,
\end{align*}
where since  $\tilde \bX_{\ma^*}(\tilde \bX_{\ma^*}^{\top}  \tilde \bX_{\ma^*})^{-1}\tilde \bX_{\ma^*}^{\top} $ is a projection matrix,
the first inequality comes from Theorem 2.22 of Puntanen (2011), and the second inequality comes from  triangle inequality.

By the definition of  $\mathbf \omega_i$, with probability 1, we have
\begin{align*}
&\max_j \|\frac{1}{\sqrt{n}}(\tilde \bX^{(j)})^{\top}
\tilde {\mathbf x}_\omega\|\leq \sum_{j=1}^{q_n+1} \|\tilde \bX_{(j)}^{\top}   \tilde {\mathbf x}_{\omega,(j)} \|\\
&\leqslant\sum_{\ell=1}^{s}\left\|\frac{1}{\sqrt{n}}\sum_{i\in \mathcal{I}_{k_{\ell,n}}^{(1)}}\tilde {\mathbf X}_i\tilde {\mathbf X}_i^{\top}  (-\alpha_{\ell})\mathbf \delta_{\ell}^*+
\frac{1}{\sqrt{n}}\sum_{i\in \mathcal{I}_{k_{\ell,n}}^{(2)}}\tilde {\mathbf X}_i\tilde {\mathbf X}_i^T(1-\alpha_{\ell})\mathbf \delta_{\ell}^*
\right\|\nonumber\\
&\leqslant\sum_{\ell=1}^{s}\|\mathbf \delta_\ell^*\|\|\frac{1}{\sqrt{n}}\tilde \bX_{(k_{\ell,n})}^{\top}  \tilde \bX_{(k_{\ell,n})}\|
\leqslant \frac{s \max_\ell \sharp \mathcal{I}_{k_{\ell,n}}}{\sqrt{n}}\|\Sigma_0 \|\max_{1\leqslant \ell\leqslant s}
\left\|\mathbf d_{\ell}^*\right\|.
\end{align*}

By (A4),  we have $\frac{\max_\ell \sharp \mathcal{I}_{k_{\ell,n}}}{\sqrt{n}}=O(1)$. Thus $\max_j \|\frac{1}{\sqrt{n}}(\tilde \bX^{(j)})^{\top}
\tilde {\mathbf x}_\omega\|$ is bounded with probability 1.

Assume  $\mathbf \varphi_j(\mathbf X,Y)=\mathbf X^{\top}  (Y-\mathbf X^{\top}  \beta_j^*)$.
By (A6), $\mathbf \varphi_j=0$ for all $Y \in (U_{F_j}^*+U_{E_j},U_{F_j}]$ which is the condition  (2.3) of Stute (1995). From (A1), we have $E \| \mathbf \varphi_j\|^2 <\infty$. Denote $\tilde {\mathbf X}_{t}=(\tilde x_{t,1},\ldots,\tilde x_{t,p})^{\top}  $.
By (2.5) in Stute (1995).  we have
\begin{align}\label{rate}
\frac{1}{\sqrt{\sharp\mathcal{I}_j}}\sum_{t\in \mathcal{I}_j}\tilde x_{t,i}\tilde \ep_t=\frac{1}{\sqrt{\sharp\mathcal{I}_j}}\sum_{t\in \mathcal{I}_j}U_{j,i,t}+R_{j,i},
\end{align}
where the $U_{j,i,t},t\in \mathcal{I}_j$ are i.i.d. with mean zero and $V_{j,i}=Var(U_{j,i,t})<\infty$, and
$$|R_{j,i}|=O_{a.s.}\left((\sharp\mathcal{I}_j)^{-1/2}\log(\sharp\mathcal{I}_j)\right).$$ The expression of $V_{j,i}$ can be found in Theorem 1.1 of Stute (1996).

Because there are $s+1$ difference models in each of which $T_\ell$ has the distribution $F_{\ell},\ell=1,\ldots,s+1$, and by
$a_{\ell}\in\mathcal{I}_{k_{\ell,n}}, \ell=1,\ldots,s$, for all $V_{j,i}$, $j\in [k_{\ell-1,n}+1,k_{\ell,n}-1]$  have the same value and set $V_{j,i}=V_{\ell,i}^*$ for all $j\in [k_{\ell-1,n}+1,k_{\ell,n}-1]$, $\ell=1,\ldots,s+1$, where $k_{0,n}=0,k_{s+1,n}=q_n+2$.
Assume $\sharp\mathcal{I}_{k_{\ell,n}}^{(1)}/\sharp\mathcal{I}_{k_{\ell,n}}\rightarrow \alpha_{\ell}$,  and then we can consider that $\{t_{i},i\in \mathcal{I}_{k_{\ell,n}}\}$ are independent samples with the same mixture distribution as $\tilde F_\ell=\alpha_{\ell} F_{\ell}+(1-\alpha_{\ell})F_{\ell+1}$. Thus when $j=k_{\ell,n}$, $V_{j,i}=\tilde V_{\ell,i}=\alpha_{\ell}V_{\ell,i}^*+(1-\alpha_{\ell})V_{\ell+1,i}^*$,
$\ell=1,\ldots,s$.
Therefore $V_{j,i},j =1,\ldots, q_n+1,i=1,\ldots,p$ have at most $(2s+1)p$ different values and we denote the greatest value among them by $V$.

By (\ref{rate}), we get
\begin{align*}
&\max_{1\leq j\leq q_n+1}\|\frac{1}{\sqrt{n}}(\tilde \bX^{(j)})^{\top}  \tilde {\mathbf \ep}\|\\
&\leq \sum_{i=1}^p \max_{1\leqslant k\leqslant q_n+1}
|\frac{1}{\sqrt{n}}\sum_{j=k}^{q_n+1}\sum_{t\in \mathcal{I}_j}\tilde x_{t,i}\tilde \ep_t|\\
&\leq\sum_{i=1}^p \max_{1\leqslant k\leqslant q_n+1}
|\frac{1}{\sqrt{n}}\sum_{j=k}^{q_n+1}\sum_{t\in \mathcal{I}_j}
U_{j,i,t}|+\sum_{i=1}^p\sum_{j=1}^{q_n+1}\frac{\sqrt{\sharp \mathcal{I}_j}}{\sqrt{n}}|R_{j,i}|
\end{align*}

For $n^*=O(n), q_n=O(m)=O(n^{1/2})$,  we have $\sum_{i=1}^p\sum_{j=1}^{q_n+1}\frac{\sqrt{\sharp \mathcal{I}_j}}{\sqrt{n}}|R_{j,i}|=O_{a.s.}(\log n)$.

By H\'{a}jek and R\'{e}nyi inequality (H\'{a}jek and R\'{e}nyi, 1955), for any given $\epsilon>0$, we have
\begin{align*}
&\sum_{n=2}^{\infty}\frac{1}{n}P(\max_{1\leqslant k\leqslant q_n+1}
|\frac{1}{\sqrt{n}}\sum_{j=k}^{q_n+1}\sum_{t\in \mathcal{I}_j}
U_{j,i,t}|\geq \epsilon \log n)\\
&\leq \sum_{n=2}^{\infty}\frac{1}{\epsilon ^2 n (\log n)^2} \sum_{j=1}^{q_n+1}\frac{\sharp \mathcal{I}_j}{n} V_{j,i}<
\sum_{n=2}^{\infty}\frac{V}{\epsilon ^2 n(\log n)^2}< \infty .
\end{align*}
By Theorem 2.1 of Yang, Su and Yu (2008), we have  $$\sum_{i=1}^p \max_{1\leqslant k\leqslant q_n+1}
|\frac{1}{\sqrt{n}}\sum_{j=k}^{q_n+1}\sum_{t\in  \mathcal{I}_j}
U_{j,i,t}|=o_{a.s.}(\log n).$$

Combining the above result, we have
$\frac{1}{\sqrt{n}}\max_{j\not\in \ma}||(\tilde \bX^{(j)})^{\top}  (\tilde
{\mathbf y}-\tilde \bX_{\ma^*}\hat{\mathbf \theta}_{\ma^*})||=O_{a.s.}(\log n)$, but $\sqrt{n}\lambda_n/\log n\to \infty$,
 and then (\ref{KKT1}) holds with probability 1.

Let $\ma^*=\{j_i,i=1,\ldots,\hat b\}$ where $j_1<j_2<\ldots<j_{\hat b}$. By the condition (A7) and (A8), we have $j_1=1$ and $\hat b \leq 2s+1$.
Put $\tilde \bX_{(c:d)}=(\tilde \bX_{(c)}^{\top}  ,\tilde \bX_{(c+1)}^{\top}  ,\ldots,\tilde \bX_{ (d)}^{\top}  )^{\top}  $  and $\tilde {\mathbf y}_{(c:d)}=(\tilde {\mathbf y}_{(c)}^{\top}  ,\tilde {\mathbf y}_{(c+1)}^{\top}  ,\ldots,\tilde {\mathbf y}_{ (d)}^{\top}  )^{\top}  $ for $c\leq d$,
 and $\tilde {\mathbb{X}}_{i}=\tilde \bX_{(j_{i}:j_{i+1}-1)}^{\top}   \tilde \bX_{((j_{i}:j_{i+1}-1)}=\sum_{k=j_i}^{j_{i+1}-1}
 \tilde \bX_{(k)}^{\top}  \tilde \bX_{(k)}$, $i=1,\ldots,\hat b$, $j_{\hat b+1}=q_n+2$. Then we have
\begin{align*}
&\left(\tilde \bX_{\ma^*}^{\top}  \tilde \bX_{\ma^*}\right)^{-1}\\&=\begin{pmatrix}
\tilde {\mathbb{X}}_{1}^{-1}&-\tilde {\mathbb{X}}_{1}^{-1}&0&\ldots&0&0&0\\
-\tilde {\mathbb{X}}_{1}^{-1}&\tilde {\mathbb{X}}_{1}^{-1}+\tilde {\mathbb{X}}_{2}^{-1}&-\tilde {\mathbb{X}}_{2}^{-1}&\ldots&0&0&0\\
\vdots&\vdots&\vdots&\vdots&\vdots&\vdots&\vdots\\
0&0&0&\ldots&-\tilde {\mathbb{X}}_{\hat b-2}^{-1}&\tilde {\mathbb{X}}_{\hat b-2}^{-1}+\tilde {\mathbb{X}}_{\hat b-1}^{-1}&-\tilde {\mathbb{X}}_{b-1}^{-1}\\
0&0&0&\ldots&0&-\tilde {\mathbb{X}}_{\hat b-1}^{-1}&\tilde {\mathbb{X}}_{\hat b-1}^{-1}+\tilde {\mathbb{X}}_{\hat b}^{-1}&\\
\end{pmatrix}.
\end{align*}
Thus
{\small\begin{align}\label{theta}
\hat {\mathbf \theta}_{\ma^*}&=\left(\tilde \bX_{\ma^*}^{\top}  \tilde \bX_{\ma^*}\right)^{-1}\tilde \bX_{\ma^*}^{\top}  \tilde {\mathbf y}_{\ma^*}\nonumber\\&=
\begin{pmatrix}\tilde {\mathbb{X}}_{1}^{-1}\tilde \bX_{(j_{1}:j_{2}-1)}^{\top}  \tilde {\mathbf y}_{(j_{1}:j_{2}-1)}\\
\tilde {\mathbb{X}}_{2}^{-1}\tilde \bX_{(j_{2}:j_{3}-1)}^{\top}  \tilde {\mathbf y}_{(j_{2}:j_{3}-1)}-\tilde {\mathbb{X}}_{1}^{-1}\tilde \bX_{(j_{1}:j_{2}-1)}^{\top}  \tilde {\mathbf y}_{(j_{1}:j_{2}-1)}\\
\vdots\\
\tilde {\mathbb{X}}_{\hat b}^{-1}\tilde \bX_{(j_{\hat b}:j_{\hat b+1}-1)}^{\top}  \tilde {\mathbf y}_{(j_{\hat b}:j_{\hat b+1}-1)}-\tilde {\mathbb{X}}_{\hat b -1}^{-1}\tilde \bX_{(j_{\hat b-1}:j_{\hat b}-1)}^{\top}  \tilde {\mathbf y}_{(j_{\hat b-1}:j_{\hat b}-1)}
\end{pmatrix}.
\end{align}
}
By (\ref{cp}), $m=\lfloor c\sqrt{n^*}\rfloor$ and (\ref{theta}), we have the following results:
\begin{itemize}
\item[1.] $\hat {\mathbf \theta}_1\to_{a.s.} \mathbf \beta_1^*$,
\item[2.] If $k_{j,n}\in \ma^*$ and $k_{j+1,n}\not\in \ma^*$, $\hat {\mathbf \theta}_{k_{j,n}}\to_{a.s.} \mathbf d_j^*$,
\item[3.] If $k_{j,n}\not\in \ma^*$ and $k_{j+1,n}\in \ma^*$, $\hat {\mathbf \theta}_{k_{j+1,n}}\to_{a.s.} \mathbf d_j^*$,
\item[4.] If $k_{j,n}\in \ma^*$ and $k_{j+1,n}\in \ma^*$, $\hat {\mathbf \theta}_{k_{j,n}}+\hat {\mathbf \theta}_{k_{j+1,n}}\to_{a.s.} \mathbf d_j^*$.
\end{itemize}
By { assuming} $\min\{||\beta_1||_2,||\delta_j||_2\}>2\sqrt{p} \gamma\lambda$, (\ref{KKT2}) holds with probability 1. Hence the proof of Theorem \ref{thm2}
is complete.

\subsection{Proof of Theorem \ref{thm3}}\label{app4}

{ Following the definition of $\hat s$ and the condition (\ref{cona}), the result (1) can be directly proved by Theorem \ref{thm2}.
Based on the condition (\ref{cona}), the result (2) can be proved by applying Theorem \ref{thm1} to each single change point in disjoint intervals.}
Under the Assumption (A7) and the result (2), $\tilde \bX_{\mathbf {\hat a}}^T\tilde \bX_{\mathbf {\hat a}}/n\to_{a.s.} \Gamma>0$. Thus the criterion $$\frac{1}{2n}\| \tilde {\mathbf y}_{\mathbf {\hat a}}- \tilde \bX_{\mathbf {\hat a}} \mathbf \theta \|^2+ \sum_{i=1}^{( s+1)p}
p_{\lambda_n,\gamma}(|\theta_{i}|)$$ is strictly convex. By the KKT condition  and result (2), (3) holds  with probability 1 in the following events
\begin{align*}
\max_{j\not\in S}|n^{-1}(\tilde {\mathbf X}_j)^{\top}  (\tilde {\mathbf y}_{\mathbf {\hat a}}- \tilde \bX_{\mathbf {\hat a}} \hat{\mathbf \theta}^o  )|\leq \lambda_n~\textrm{and}~
\min_{j\in S}\{\|\hat{\theta}_j^o\|\}\geq\gamma\lambda_n.
\end{align*}
which can be proven similarly as the proof of Theorem \ref{thm2}. Hence the proof is complete.

%\section*{Acknowledgements}
%And this is an acknowledgements section with a heading that was produced by the
%$\backslash$section* command. Thank you all for helping me writing this
%\LaTeX\ sample file. See \ref{suppA} for the supplementary material example.

%\begin{supplement}
%\sname{Supplement A}\label{suppA}
%\stitle{Title of the Supplement A}
%\slink[url]{http://www.e-publications.org/ims/support/dowload/imsart-ims.zip}
%\sdescription{Dum esset rex in
%accubitu suo, nardus mea dedit odorem suavitatis. Quoniam confortavit
%seras portarum tuarum, benedixit filiis tuis in te. Qui posuit fines tuos}
%\end{supplement}


\begin{thebibliography}{9}


\bibitem[Bai(2003]{bai03} Bai, J.  and Perron, P. (2003), Computation and Analysis of Multiple Structural Change Models, \emph{Journal of Applied Econometric}, 18, 1-22.

\bibitem[Buckkey and Jammes(1979)]{buck1979} Buckley, J. and James, I. (1979), Linear regression with censored data, \emph{Biometrika}, 66, 429-436.

\bibitem[Dave et.al. (2004)]{dave04} Dave, S. S., Wright, G., Tan, B., Rosenwald, A., Gascoyne, R. D., Chan, W. C., $\ldots$  Staudt, L. M. (2004), Prediction of survival in follicular lymphoma based on molecular features of tumor-infiltrating immune cells, \emph{New England Journal of Medicine}, 351, 2159-2169.

\bibitem[Davis et.al. (2006)]{davis06}Davis, R. A., Lee, T. C. M., Rodriguez-Yam, G. A. (2006), Structural break estimation for nonstationary
time series models, \emph{Journal of the American Statistical Association}, 101, 223-239.



\bibitem[Fan(2001)]{fan01} Fan, J. and Li, R. (2001), Variable selection via nonconcave penalized likelihood and its
oracle properties, \emph{Journal of the American Statistical Association},  96,
1348-1360.

\bibitem[Fearnhead(2009)]{fearnhead09} Fearnhead, P.  and Vasileiou, D. (2009), Bayesian Analysis of
Isochores, \emph{Journal of the American Statistical Association}, 104, 132-141.

\bibitem[Gordon(1981)]{gordon1981} Gordon, A. D. (1981), {\em Classification}. London: Chapman and Hall.

\bibitem[H\'{a}jek and R\'{e}nyi(1955)]{HR1955}
 H\'{a}jek, J. and R\'{e}nyi, A. (1955), Generalization of an inequality of Kolmogorov,
\emph{Acta Mathematica Academiae Scientiarum Hungarica}, 6, 281-283.

\bibitem[Hansen,(2000)]{Hansen00}Hansen, B. E. (2000), Sample splitting and threshold estimation,  \emph{Econometrica},68, 575-603.

\bibitem[Harchaoui and {L\'-Leduc} (2010)]{HL10}Harchaoui, Z. and {L\'evy-Leduc}, C. (2010), Multiple Change-Point Estimation With
a Total Variation Penalty, \emph{Journal of the American
Statistical Association}, 105, 1480-1493.



\bibitem[Huang, Ma and Xie(2006)]{HMX2006}  Huang, J., Ma, S. and Xie, H. (2006), Regularized estimation in the accelerated failure time model with high-dimensional covariates, \emph{Biometrics}, 62, 813-820.



\bibitem[Inclan and Tiao(1994)]{IT1994} Inclan, C., and Tiao, G. C. (1994), Use of cumulative sums of squares for retrospective detection of changes of variance , \emph{Journal of the American Statistical Association}, 89, 913-923.


\bibitem[Jin, Shi and Wu (2013)]{JSW13} Jin, B., Shi, X. and Wu, Y. (2013), A novel and fast methodology for simultaneous multiple structural break estimation and variable selection for nonstationary time series models, \emph{Statistics and Computing}, 23,221-231.

\bibitem[Kalbfleisch and Prentice (2002)]{kalb2003} Kalbfleisch, J. D. and Prentice, R. L. (2002). {\em The Statistical Analysis of Failure Time Data}. New York: Wiley.

\bibitem[Kosorok and Song (2007)]{KS07} Kosorok, M. R. and Song, R.(2007), Inference under right censoring for transformation models with a change-point based on a covariate threshold, \emph{The Annals of Statistics}, 35, 957-989.

\bibitem[Lawless (2011)]{L2011} Lawless, J. F. (2011), \emph{Statistical models and methods for lifetime data}, John Wiley\& Sons.


\bibitem[Lin, Wei and Ying(1998)]{lin1998} Lin, D. Y., Wei, L. J. and Ying, Z. (1998), Accelerated failure time models for counting processes, \emph{Biometrika}, 85, 605-618.

\bibitem[Lou, Turnbull and Clark(1997)]{luo1997} Luo X., Turnbull, B. W. and Clark, L. C.(1997), Likelihood ratio tests for a changepoint with survival data, \emph{Biometrika}, 84, 555-565.

\bibitem[Perron(2006)]{P2006} Perron, P. (2006), \emph{Dealing with Structural Breaks, in Palgrave Handbook of Econometrics},
Vol. 1: Econometric Theory, K. Patterson and T.C. Mills (eds.), Palgrave
Macmillan, 278-352.

\bibitem[Pons (2003)]{P2003}  Pons, O.(2003), Estimation in a Cox regression model with a change-point according to a threshold in a covariate, \emph{The Annals of Statistics}, 31, 442-463.

\bibitem[Prentice(1978)]{pren1978} Prentice, R. L. (1978), Linear rank test with right censored data, \emph{Biometrika}, 65, 167-179.


\bibitem[Puntanen(2011)]{Puntanen2011} Puntanen, S. (2011), Projection Matrices, Generalized Inverse Matrices, and Singular Value Decomposition by Haruo Yanai, Kei Takeuchi, Yoshio Takane. \emph{International Statistical Review}, 79, 503-504.

\bibitem[Stute (1993)]{Stute1993} Stute, W. (1993), Consistent estimation under random censorship when covariables are present, \emph{Journal of Multivariate Analysis}, 45,89-103.


\bibitem[Stute (1995)]{Stute1995} Stute, W. (1995), The central limit theorem under random censorship, \emph{The Annals of Statistics}, 23,422-439.


\bibitem[Stute(1996)]{Stute1996} Stute, W. (1996), Distributional convergence under random censorship when covariables are present, \emph{Scandinavian Journal of Statistics}, 23,461-471.


\bibitem[Tibshirani(1996)]{tibshirani96} Tibshirani, R. (1996), Regression Shrinkage and Selection
via the lasso, \emph{Journal of the Royal Statistical Society}, Series
B, 58, 267-288.

\bibitem[Tong(2012)]{Tong12} Tong, H. (2012), \emph{Threshold models in non-linear time series analysis.}  Springer Science \& Business Media.



\bibitem[Tsiatis(1990)]{tsiatis1990} Tsiatis, A. A. (1990), Estimating regression parameters using linear rank tests for censored data, \emph{The Annals of  Statistics}, 18, 354-372.

\bibitem[Xia (2016)]{xia2016} Xia, X., Jiang, B., Li, J., Zhang, W. (2016). Low-dimensional Confounder Adjustment and High-dimensional Penalized Estimation for Survival Analysis. {\em Lifetime Data Analysis}, 22, 547-569.

\bibitem[Yang, Su and YU(2008)]{YSY2008} Yang, S., Su, C. and Yu, K. (2008), A general method to the strong law of lage numbers and its applications, \emph{Statistics and Probability Letters}, 78, 794-803.


\bibitem[Yao(1989)]{yao89} Yao, Y., and Au, S. T. (1989),  Least-Squares Estimation of
a Step Function, \emph{Sankhy\={o}: The Indian Journal of
Statistics}, Series  A, \textbf{51}, 370-381.


\bibitem[Ying(1993)]{ying1993} Ying, Z. (1993), A large sample study of rank estimation for censored regression data, \emph{The Annals of Statistics}, 21, 76-99.


 \bibitem[Yu,et.al.(2012)]{Yu2012}  Yu, T., Li, J. and Ma, S. (2012),  Adjusting confounders in ranking biomarkers:amodel-based ROC approach,  \emph{Briefings In Bioinformatics}, 13, 513-523.

%\bibitem[Ciuperca(2014)]{Ciuperca14} Ciuperca, G. (2014). Model selection by lasso methods in a change-point model. \emph{Statistical Papers},\textbf{55},349-374.










\bibitem[Zhang(2010)]{zhang10}Zhang, C. (2010), Nearly unbiased variable selection under minimax concave penalty, \emph{The Annals of Statistics}, 38, 894-942.

%
%\bibitem[Zhao(2006)]{zhao06} Zhao, P., and Yu, B. (2006) On Model
%Selection Consistency of lasso. \emph{Journal of Machine Learning
%Research}, \textbf{7}, 2541-2567.



\end{thebibliography}
\end{document}